%% file: weyl100116.tex
\documentclass[11pt]{article}
\pdfoutput=1  
\usepackage{graphicx,color}
\usepackage{appendix}
\usepackage{latexsym,amsmath,amssymb,graphicx,booktabs}
\usepackage{epsfig,latexsym,cite}
\usepackage{hyperref}
\numberwithin{equation}{section}

\definecolor{MyBlue}{rgb}{0.15,0.15,0.70}

\hypersetup{
colorlinks=true,
citecolor=MyBlue,
linkcolor=MyBlue,
urlcolor=MyBlue
}

\input epsf
\usepackage{amssymb}
\usepackage{amsmath}
\usepackage{amsfonts}
\usepackage{upgreek}
\usepackage{latexsym}

\newcommand{\LCDM}{$\Lambda$\rm{CDM}}

\include{mydefs}

\graphicspath{ {./Graphs/} }
\begin{document}

\begin{titlepage}

\vspace*{2cm}

\centerline{\Large \bf Non-local gravity with a Weyl-square term\\}

\vskip 0.4cm
\vskip 0.7cm
\centerline{\large Giulia Cusin$^a$, Stefano Foffa$^a$, Michele Maggiore$^a$ and Michele Mancarella$^{b,c,d}$ }
\vskip 0.3cm
\centerline{\em $^a$D\'epartement de Physique Th\'eorique and Center for Astroparticle Physics,}  
\centerline{\em Universit\'e de Gen\`eve, 24 quai Ansermet, CH--1211 Gen\`eve 4, Switzerland}

\vspace{3mm}
\centerline{\em $^b$Institut de physique th\' eorique, Universit\'e  Paris Saclay}
\centerline{\em CEA, CNRS, 91191 Gif-sur-Yvette, France}
\vspace{3mm}
\centerline{\em $^c$Universit\'e Paris Sud, 15 rue George Cl\'emenceau, 91405,  Orsay, France}
\vspace{3mm}
\centerline{\em $^d$Physics Department, Theory Unit, CERN, CH-1211 Gen\`eve 23, Switzerland}

\vskip 1.9cm

\begin{abstract}
Recent work has shown that modifications of General Relativity based on the addition to the action of a non-local term $R\,\Box^{-2}R$, or on the addition to the equations of motion of a term  involving $(\gmn\iBox R)$,
produce  dynamical models of dark energy which are cosmologically viable both at the background level and at the level  of cosmological perturbations. We explore a more general class of models based on the addition to the action of terms proportional to 
$\Rmn\,\Box^{-2}\RMN$ and $C_{\mu\nu\rho\sigma}\, \Box^{-2}C^{\mu\nu\rho\sigma}$, where $C_{\mu\nu\rho\sigma}$ is the Weyl tensor. We find that the term $\Rmn\,\Box^{-2}\RMN$ does not give a viable background evolution. The non-local Weyl-square term, in contrast, does not contribute to the background evolution but we find that, at the level of cosmological perturbations, it gives instabilities in the tensor sector. Thus, only non-local terms which  depend just on the Ricci scalar $R$   appear to be cosmologically viable. We discuss how these results can provide a hint for the mechanism that might generate  these effective non-local terms from a fundamental local theory.

\end{abstract}


\end{titlepage}

\newpage

\section{Introduction}

Locality is one of the fundamental principles of quantum field theory. Nevertheless, even in a fundamental local theory, non-localities in general appear
at an effective level. This can already happen  classically, when one integrates out some degree of freedom and obtains an effective theory for the remaining degrees of freedom. Cosmological perturbation theory provides a simple example. In this case,  integrating out the short-wavelength modes, one finds  effective non-local equations for the long-wavelength 
modes~\cite{Carroll:2013oxa}. At the quantum level, non-local terms  emerge when one computes the effective action that takes into account loops of massless or light particles.

Non-local terms featuring inverse powers of the d'Alembertian operator are relevant in the infrared (IR), and it is natural to ask  whether such non-localities could play a role as IR modifications of general relativity (GR), and in particular whether they could provide a dynamical explanation for dark energy. An attempt in this direction has been done in 
\cite{Deser:2007jk,Deser:2013uya,Woodard:2014iga}, adding to the Ricci scalar $R$  in the Einstein-Hilbert action a term of the form $Rf(\iBox R)$. 
In  models of this form there is no explicit mass scale, since the function $f$ is dimensionless. 
However, this class of models has been ruled out by the comparison with cosmological observations~\cite{Dodelson:2013sma}. Indeed, after fixing the function $f(\iBox R)$ so that it gives a viable  cosmological evolution at the background level, the corresponding cosmological perturbations fail to reproduce structure-formation data. 

The situation is more promising for models where the non-local terms depend on an explicit mass scale (whose physical origin could be due to mechanisms such as those discussed recently in \cite{Maggiore:2015rma}). 
Models of this type have been explored in the context of the degravitation mechanism~\cite{ArkaniHamed:2002fu,Dvali:2006su,Dvali:2007kt}. Partly inspired by these works, as well as by
attempts at writing massive gravity in non-local form~\cite{Porrati:2002cp,Jaccard:2013gla}, in  
ref.~\cite{Maggiore:2013mea} has been proposed a phenomenological  modification of gravity, based on
the non-local equation of motion
\be\label{RT}
\Gmn -(1/3)m^2\(\gmn\iBox R\)^{\rm T}=8\pi G\,\Tmn\, ,
\ee
where the  superscript T denotes the operation of taking the transverse part of a tensor (which is itself a nonlocal operation), and $m^2$ is a mass parameter. The extraction of the transverse part ensures that energy-momentum conservation is  automatically satisfied. A closed form for the action corresponding to \eq{RT} is not known. A related non-local model, proposed in \cite{Maggiore:2014sia}, is defined by the action
\be\label{RR}
S=\frac{\mplr^2}{2}\int d^{4}x \sqrt{-g}\, 
\[R-\frac{1}{6} m^2R\frac{1}{\Box^2} R\]\, ,
\ee
where $\mplr=1/\sqrt{8\pi G}$ is the reduced Planck mass, and $m$ a mass parameter. 
These two models are related by the fact that,  linearizing over flat space, they give  the same equations of motion. However at the full non-linear level, or linearizing over a background different from Minkowski (such as FRW), they are different. Both in \eq{RT} and in the equations of motion derived from (\ref{RR}), the inverse d'Alembertian is taken to be the one defined using the retarded Green's function, to ensure causality. A non-local action such as (\ref{RR}) must indeed be understood as an effective action which takes into account quantum loop effect. For the in-out expectation values of a quantum fields the  variation of such a non-local effective action gives non-local equations of motion which depend on the Feynman propagator. However, 
the equations of motion for the in-in expectation values, which can be obtained with the Schwinger-Keldysh formalism, depend on the inverse d'Alembertian defined with the 
retarded Green's function, and are therefore causal.

As discussed in \cite{Maggiore:2013mea,Maggiore:2014sia,Foffa:2013vma,Nesseris:2014mea,Dirian:2014ara},  in the comparison with cosmological observations the  models (\ref{RT}) and (\ref{RR}) perform remarkably well.  They dynamically generate  a dark energy and have a realistic background FRW evolution. Their cosmological perturbations are well-behaved 
and their quantitative effects are consistent with CMB, supernovae, BAO and structure formation data~\cite{Nesseris:2014mea,Dirian:2014ara,Barreira:2014kra}. Implementing the cosmological perturbations of the non-local models into a Boltzmann code and performing parameter estimation and a  fit to CMB, supernovae and BAO data, one finds that these models fit very well the data, with a $\chi^2$ comparable to that of 
$\Lambda$CDM~\cite{Dirian:2014bma}.\footnote{Actually  in ref.~\cite{Diraninprep}, updating the analysis of 
\cite{Dirian:2014bma} using the 2015 {\em Planck\,} data as well as an extended BAO datasets, and performing a Bayesian model comparison, it will be shown that the model (\ref{RT}) and $\Lambda$CDM both fit equally well the data, while the model (\ref{RR}), even if by itself fits the data well, is disfavored in the Bayesian comparison with 
$\Lambda$CDM  or with the model (\ref{RT}).}
It should be stressed that both the model (\ref{RT}) and the model (\ref{RR})  have the same number of parameters as $\Lambda$CDM, with the mass $m$ replacing the cosmological constant, and in this sense they are the only existing models that can compete quantitatively with 
$\Lambda$CDM from the point of view of fitting the data, without being merely an extension of $\Lambda$CDM with extra free parameters.
Further work on these models has been presented in \cite{Modesto:2013jea,Foffa:2013sma,Kehagias:2014sda,Conroy:2014eja,Cusin:2014zoa,Dirian:2014xoa,Mitsou:2015yfa,Barreira:2015fpa,Barreira:2015vra}.

The purpose of this paper is to investigate whether some possible generalizations of these non-local theories
are cosmologically viable. We are not driven here by the aim of improving the agreement with the cosmological observations. From this point of view the one-parameter models (\ref{RT}) and (\ref{RR}) already work so well that, with present data, there is not much point in enlarging the parameter space. Rather, our question is of more conceptual nature. Indeed, the models (\ref{RT}) and (\ref{RR}) have been proposed on a purely phenomenological ground. The next  step should be to understand how they could be generated from a fundamental local theory. From this point of view,  it is important 
to understand whether the fundamental local theory should generate exactly a non-local structure such as that given in
\eq{RT} or in \eq{RR}, or whether  more general non-local structures are phenomenologically acceptable. This would give a precious hint on the mechanism at work for the generation of the  non-local terms. In this sense, the results of the present paper are meant to pave the way for future works aiming at understanding the origin of the non-local terms.

In this paper we study  the generalization of non-local actions of the type (\ref{RR}), at the level quadratic in curvatures.
As a basis for the curvature-square terms one could use $\Rmnrs^2$, $\Rmn^2$ and $R^2$. However, for cosmological application it is more convenient to trade the Riemann tensor $\Rmnrs$ for the Weyl tensor $C_{\rho\sigma\mu\nu}$, which in $d=3$ spatial dimensions is given by
\be \label{weyl}
C_{\mu\nu\rho\sigma}=R_{\mu\nu\rho\sigma}- \left( g_{\mu[ \rho} R_{\sigma ] \nu}-g_{\nu [ \rho} R_{\sigma ] \mu}\right)+ \frac{1}{3} g_{\mu [ \rho} g_{\sigma ] \nu} R\, ,
\ee
where $[\mu\nu]$ denotes the anti-symmetrization over the indices, e.g. $T_{[\mu\nu]}=(1/2)(T_{\mu\nu}-T_{\nu\mu})$.
We will then consider an action of the form
\be\label{actionTotal}
S_{\rm NL}=\frac{\mplr^2}{2}\int d^4 x \sqrt{-g}\,
\left[R-\mu_1 R\frac{1}{\Box^2}R-\mu_2 C^{\mu\nu\rho\sigma}\frac{1}{\Box^2}C_{\mu\nu\rho\sigma}-\mu_3\RMN\frac{1}{\Box^2}\Rmn
\right]\,,
\ee
where $\mu_1$, $\mu_2$ and $\mu_3$ are  independent parameters with dimension of squared mass. Another notable combination is the one that corresponds to the Gauss-Bonnet term, 
\be
E=\Rmnrs^2-4\Rmn^2+R^2\, , 
\ee
so instead of trading
 $\Rmnrs\Box^{-2}\RMNRS$ for  $C^{\mu\nu\rho\sigma}\Box^{-2}C_{\mu\nu\rho\sigma}$ one could trade it 
for the combination
$\Rmnrs\Box^{-2}\RMNRS -4  \Rmn\Box^{-2}\RMN
+R\Box^{-2}R$. However, the choice of basis in \eq{actionTotal} is more convenient for cosmological applications, because the Weyl tensor vanishes in FRW.  

It should  be made clear that the space of all possible non-local theories is of course very large, and our choice by no means exhaust all possibilities. For instance,  taking the transverse part of a tensor opens up new possibilities for writing non-local equations of motions, such as \eq{RT}, which are not easily expressed  in terms of an action. However, while the addition to Einstein equations of a term proportional to $\(\gmn\iBox R\)^{\rm T}$ gives a viable cosmological model, 
the addition of term proportional to $(\iBox\Rmn)^T$ has been shown to be cosmologically non-viable, already at the background 
level~\cite{Maggiore:2013mea,Foffa:2013vma}.
Several other options are possible, which can be considered as non-linear extensions of \eq{RR} or of 
\eq{actionTotal}. A particularly well-motived extension of \eq{RR} is obtained writing~\cite{Cusininprep}
\be\label{SDelta4}
S=\frac{\mplr^2}{2}\int d^{4}x \sqrt{-g}\, 
\[R-\frac{1}{6} m^2R\frac{1}{\Delta_4} R\]\, ,
\ee
where
\be\label{Delta4}
\Delta_4=\Box^2+2\RMN\n_{\mu}\n_{\nu}-\frac{2}{3} R\Box +\frac{1}{3}(\n^{\mu}R)\n_{\mu}
\, ,
\ee
is the so-called Paneitz operator.
Indeed, this operator enters in the computation of the anomaly-induced effective action which, as discussed in
\cite{Maggiore:2015rma}, and as we will recall in the Conclusions, could be at the origin of these effective non-local IR modifications of GR. More generally, we could also replace $\Box^2$ by $\Delta_4$ in \eq{actionTotal}.

In the present paper we study the cosmological consequences of the action (\ref {actionTotal}), both at the level of background evolution and at the level of cosmological perturbations. The plan of the paper is as follows.
In Sect.~\ref{sect:Rmnterms} we discuss the effect of the term $\Rmn\,\Box^{-2}\RMN$, and we will see that it is ruled out already at the background level. We then turn to the effect of adding  the non-local Weyl  term 
$C^{\mu\nu\rho\sigma}\Box^{-2}C_{\mu\nu\rho\sigma}$ to the theory that already features the 
$R\Box^{-2}R$ term. In Sect.~\ref{sect:Weyl} we show how to write the non-local Weyl  term in local form introducing auxiliary fields. The cosmological consequences of the model are   studied in Sect.~\ref{cosmology}. The non-local Weyl  term does not affect the background evolution, but only the cosmological perturbations. The background evolution is therefore the same induced by the $R\Box^{-2}R$ term, which is briefly reviewed in Sect.~\ref{sect:back}.
The decomposition into  scalars, vectors and tensors  of the perturbations associated to non-local Weyl  term  is non-trivial, and we discuss it in Sect.~\ref{sect:deco}.
Scalar perturbations are then discussed in Sect.~\ref{sect:scal} and tensor perturbations in  Sect.~\ref{sect:tenssect}.
We draw our conclusions in Sect.~\ref{sect:concl}. In the appendixes we perform a more detailed comparison between analytic and numerical solutions of the equations in the scalar and tensor sector.

We use  the MTW conventions for the metric, Riemann tensor, etc., so in particular our  signature is $(-, +, +, +)$, and
we set $\hbar=c=1$. A prime will denote the derivative with respect to $x\equiv \log a$, where $a$ is the scale factor in FRW.

\section{The $\Rmn\,\Box^{-2}\RMN$ term}\label{sect:Rmnterms}

The term proportional to $\mu_2$ in the action (\ref{actionTotal}) is quadratic in the Weyl tensor. Thus,  in the equations of motion it gives a contribution linear in $C_{\mu\nu\rho\sigma}$, which vanishes in FRW. Therefore this term does not contribute to the background evolution, but only at the level of  cosmological perturbation. To study first of all whether the model (\ref{actionTotal})
is viable at the level of background evolution, we then only need to consider the action (\ref{actionTotal})  with $\mu_2=0$.
The corresponding equations of motion can be obtained in a standard way 
localizing the action through the introduction of   auxiliary scalar fields, see e.g.
 \cite{Nojiri:2007uq,Jhingan:2008ym,Koshelev:2008ie,Koivisto:2009jn,Barvinsky:2011rk,Deser:2013uya,Jaccard:2013gla,Maggiore:2013mea,Maggiore:2014sia}. In particular, 
we introduce two auxiliary scalar fields 
\be\label{auxscalfields}
U\equiv-\Box^{-1} R\, ,\qquad\qquad  S\equiv-\Box^{-1} U=\,\Box^{-2}\,R ,
\ee
and two auxiliary tensor fields, 
\be \label{tensfields}
U_{\mu\nu}\equiv-\Box^{-1}\,R_{\mu\nu}\, , 	
\qquad S_{\mu\nu}\equiv-\Box^{-1}U_{\mu\nu}=\Box^{-2}\,R_{\mu\nu}\, .
\ee
These relations can be implemented at the level of the action introducing four Lagrange multipliers
$\zeta_1,\zeta_2,\zeta^{\mu\nu}_1$ and $\zeta^{\mu\nu}_2$, and writing
\bees\label{localizedactionRmn}
S_{\rm NL}&=&\frac{\mplr^2}{2}\int d^4x \sqrt{-g}
\[ R \left(1-\mu_1 S\right)-\mu_3\Rmn S^{\mu\nu}
-\zeta_1\left(\Box U+R\right)-\zeta_2\left(\Box S+U\right)\right.\nn\\
&&\hspace*{26mm}\left.
-\zeta^{\mu\nu}_1\left(\Box U_{\mu\nu}+R_{\mu\nu}\right)
-\zeta^{\mu\nu}_2\left(\Box S_{\mu\nu}+U_{\mu\nu}\right)
\]\, .
\ees
The variation with respect to the  Lagrange multipliers and with respect to the auxiliary fields enforces the constraints
$\zeta_1=\mu_1 S$, $ \zeta_2=\mu_1 U$, 
$\zeta^{\mu\nu}_1=\mu_3 S^{\mu\nu}$ and
$\zeta_2^{\mu\nu}=\mu_3 U^{\mu\nu}$.
Varying the action (\ref{localizedactionRmn}) with respect to the metric, substituting these constraints and making use of \eqs{auxscalfields}{tensfields} afterwards, we find the following covariant equation
\be \label{fieldeqs}
G_{\mu\nu}-\mu_1 \mathcal{K}_{\mu\nu}+\mu_3 \mathcal{I}_{\mu\nu}=8 \pi G T_{\mu\nu}\, ,
\ee
where
\be\label{K}
\mathcal{K}_{\mu\nu}=2S\Gmn-2\n_{\mu}\pan S -2U\gmn+\gmn\parho S\paR U
 -(1/2)\gmn U^2-(\pam S\pan U+\pan S\pam U)\, ,
\ee
is the part coming from the $R\Box^{-2}R$ term, already  found in \cite{Maggiore:2014sia}, and
\bees
\mathcal{I}_{\mu\nu}&=&\Big \{-2 S_{ \mu \alpha} G_{ \nu} ^{\alpha}- R S_{\mu\nu}-\frac{1}{2} \Box S_{\mu\nu}  -U_{\mu }^{\alpha}U_{\nu \alpha} \nonumber\\
  &+&\frac{1}{2} g_{\mu\nu} \Big( G^{\alpha \beta} S_{\alpha \beta}+\frac{1}{2}U_{\alpha \beta} U^{\alpha \beta}+\frac{1}{2} R S^{\alpha}_{\;\alpha}-\n^{\rho}S^{\alpha \beta} \n_{\rho}U_{\alpha \beta}-\n_{ \alpha}\n_{ \beta}S^{\alpha \beta}\Big) \nonumber\\
  &+&\Big[\frac{1}{2}\n_{ \mu}U_{\alpha \beta} \n_{ \nu}S^{\alpha \beta}+ \n_{ \beta}U_{ \alpha} ^{\beta}\n_{ \mu}S_{ \nu} ^{\alpha}-S_{ \nu} ^{\alpha} \n_{ \beta}\n_{ \mu}U_{ \alpha} ^{\beta}-U_{ \nu} ^{\alpha} \Box S_{ \mu \alpha}  \nonumber\\
&+&   U^{\alpha \beta}   \n_{ \beta}\n_{ \mu}S_{ \nu \alpha}-\n^{ \beta}S_{ \mu} ^{\alpha} \n_{ \nu}U_{\alpha \beta}+(U \leftrightarrow S) \Big]\Big \}+\Big \{ \mu \leftrightarrow \nu \Big \} \; .
 \ees
We now specialize to a flat FRW metric,
$ds^2=-dt^2+a^2(t)d\vx^2$ in $d=3$ spatial dimensions and in the presence of an energy-momentum tensor $T_{\mu \nu}=(\rho, a^2(t) \,p\, \delta_{ij})$. 
Then,  only the $(0,0)$ and the $(i,i)$ component of the field equations are non-trivial. It is convenient to 
introduce the variables $U=U^0_0+U^i_i$, $V=U^0_0-U^i_i/3$ (where the sum over $i=1,2,3$ is understood). Then the non-trivial equations are
\begin{eqnarray}\label{UV}
&&\pa_t^2U+3H\pa_tU=6\pa_tH+12H^2 \, , \\
&&\pa_t^2V+3H\pa_tV-8H^2\,V=2\pa_t{H}\, ,\label{pat2V}
\end{eqnarray}
where $\pa_t$ is the derivative with respect to cosmic time. It is easy to see that the equation for $V$ contains growing modes. To this purpose it is convenient to
parametrize the time evolution in terms  
of the variable
$x\equiv \log{a}$. Then \eq{pat2V} reads
\be \label{Veqn}
V^{\prime\prime}+V^{\prime}\,\left(3+\zeta\right)-8V=-2\zeta
\ee
where $f'\equiv df/dx$ and $\zeta(x)=h'(x)/h(x)$. During RD, MD and a De~Sitter phase
$\zeta(x)$ can be approximated by a constant $\zeta_0$, with $\zeta_0=\{-2,-3/2,0\}$, respectively. 
The homogeneous equation associated to \eq{Veqn} is the same encountered in the non-local model introduced in \cite{Jaccard:2013gla} and further analyzed in \cite{Foffa:2013vma}. The solution in the  approximation $\zeta(x)=\zeta_0$ is
\begin{equation}
V(x)=\frac{\zeta_0}{4}+v_0\,e^{\beta_+ x}+v_1\,e^{\beta_- x}\, ,
\end{equation}
with 
$2\beta_{\pm}=-(3+\zeta_0) \pm \sqrt{(3+\zeta_0)^2+32}$. We see that $\beta_+ >0$ both in MD and RD, which leads to a growing solution for $V$. This affects the whole cosmological evolution, since the function $V$ is coupled to all other variables through its contribution to $\mathcal{I}_{00}$. So, similarly to the model discussed in \cite{Jaccard:2013gla} and  in app.~A of \cite{Foffa:2013vma},
based on a term $(\iBox\Gmn)^T$ in the equations of motion, there is no stable cosmological evolution. In particular, if we start close to a standard FRW solution at early times, the evolution quickly departs from it and explodes. Thus, the term $\mu_3\RMN\Box^{-2}\Rmn$ is not cosmologically viable (unless of course the parameter $\mu_3$ is tuned to be extremely suppressed with respect to $\mu_1$; imposing such a condition on a phenomenological model is quite unnatural so we do not further explore this possibility).

This result is  similar to the one found in \cite{Ferreira:2013tqn}, where it was shown that a term 
$\RMN\Box^{-1}\Rmn$ also produces instabilities in the cosmological evolution. Observe that the latter term is rather of the Deser-Woodard type, i.e. of the form $\RMN f(\Box^{-1}\Rmn)$, with a dimensionless function $f$ and no explicit mass scale $m$. However, in both cases the instability is ultimately due to the form of the $\iBox$ operator on the tensor $\Rmn$.

\section{Non-local Weyl-square term and auxiliary fields}\label{sect:Weyl}

The study of the background evolution shows that the term $\RMN\Box^{-2}\Rmn$ cannot be present in a viable model, i.e. $\mu_3=0$, but tells us nothing about the non-local Weyl-square term, since the latter does not contribute to the background evolution. To see whether  this term is viable we must move one step forward, and see if the  cosmological perturbations in the presence of this term are well-behaved. We then consider the model with  action
\be\label{action}
S_{\rm NL}=\frac{\mplr^2}{2}\int d^4 x \sqrt{-g}\,\left[R-\mu_1 R\frac{1}{\Box^2}R-\mu_2 C^{\mu\nu\alpha\beta}\frac{1}{\Box^2}C_{\mu\nu\alpha\beta}\right]\,.
\ee
The equations of motion of this model can again be derived introducing auxiliary scalar fields. In this case we need again two auxiliary scalar fields 
\be\label{loc1}
U=-\Box^{-1} R\,,\hspace{2 em}S=-\Box^{-1} U=\Box^{-2} R\, ,
\ee
as well as two auxiliary fields with the symmetry properties of the Weyl tensor, defined as
\be\label{loc2}
U_{\mu\nu\alpha\beta}=-\Box^{-1} C_{\mu\nu\alpha\beta}\,,\hspace{2 em}S_{\mu\nu\alpha\beta}=-\Box^{-1} U_{\mu\nu\alpha\beta}=\Box^{-2} C_{\mu\nu\alpha\beta}\,.
\ee
Proceeding as in sect.~\ref{sect:Rmnterms}, we 
find the following covariant equation
\be\label{coveq}
G_{\mu\nu}-\mu_1 \mathcal{K}_{\mu\nu}-2\mu_2 S_{\mu\alpha\nu\beta}R^{\beta\alpha}-2\mu_2\left(\n^{\alpha}\n^{\beta}+\n^{\beta}\n^{\alpha}\right)S_{\mu\alpha\nu\beta}-\mu_2 \mathcal{P}_{\mu\nu}=8 \pi G T_{\mu\nu}\,,
\ee
where $\mathcal{K}_{\mu\nu}$ is the same as in \eq{K}, and
\begin{align}\label{Pmunu}
\mathcal{P}_{\mu\nu}&=\Big\{  S_{\mu\sigma\alpha\beta}C_{\nu}^{\,\,\,\sigma\alpha\beta}-\frac{1}{4}g_{\mu\nu}\,C^{\alpha\beta\rho\sigma}S_{\alpha\beta\rho\sigma}+\frac{1}{4}g_{\mu\nu}\Box\left(S_{\alpha\beta\rho\sigma}U^{\alpha\beta\rho\sigma}\right)-\left(\n_{\mu}S_{\alpha\beta\rho\sigma}\right)\left(\n_{\nu}U^{\alpha\beta\rho\sigma}\right)\nn\\
&\left.-2\left[\left(\n^{\sigma}\n_{\mu}U_{\nu}^{\,\,\,\alpha\rho\sigma}\right)S_{\sigma\alpha\rho\sigma}-\left(\n_{\beta}\n_{\mu}U^{\beta\sigma\rho\alpha}\right)S_{\nu\sigma\rho\alpha}+(U \leftrightarrow S)\right]\right.
\\
&-2\left[\left(\n_{\sigma}S^{\sigma\rho\alpha\beta}\right)\left(\n_{\mu}U_{\nu\rho\alpha\beta}\right)-\left(\n_{\beta}S_{\mu\alpha\rho\sigma}\right)\left(\n_{\nu} U^{\beta\alpha\rho\sigma}\right)+(U \leftrightarrow S)\right]\Big\}+\Big\{\mu \leftrightarrow \nu\Big\}\,.\nn
\end{align}
The set of equations describing the model is therefore given by eqs.~(\ref{coveq}) and  (\ref{Pmunu}), together with 
\be\label{eq1}
\Box U=- R\,,\hspace{2 em}\Box S=- U\,,
\ee
and 
\be\label{eq1tens}
\Box U_{\mu\nu\alpha\beta}=- C_{\mu\nu\alpha\beta}\,,\hspace{2 em}
\Box S_{\mu\nu\alpha\beta}=- U_{\mu\nu\alpha\beta}\,.
\ee
As discussed in detail  in 
\cite{Koshelev:2008ie,Koivisto:2009jn,Barvinsky:2011rk,Deser:2013uya,Foffa:2013sma}, transforming a non-local equation such as $U=-\iBox R$  into the local equation $\Box U=- R$ introduces spurious solutions, given by the most general solution of the associated homogeneous equation $\Box U=0$. These spurious solutions are eliminated fixing once and for all the boundary condition of the equation $\Box U=0$ (which corresponds to choosing once and for all the definition of the $\iBox$ operator). As in  \cite{Maggiore:2013mea,Maggiore:2014sia,Foffa:2013vma,Dirian:2014bma}, we will fix the boundary conditions requiring that the auxiliary fields vanish deep in RD.\footnote{In general, if the homogeneous solutions for the auxiliary fields during RD  only have decaying modes, as is the case for the models (\ref{RT}) and (\ref{RR}), choosing the initial conditions that correspond to setting them to zero, e.g. $U_{\rm hom}=0$, 
is very natural, because this solution  is an attractor. If in contrast there are growing modes, one must first of all  check whether these instabilities are fatal for the viability of the model, and in any case the results will depends on the initial conditions chosen.}

We also observe that, since the modified Einstein equations (\ref{coveq}) are obtained varying an action invariant under diffeomorphisms, the energy-momentum tensor $T_{\mu\nu}$ is automatically covariantly conserved, 
$\n_{\mu}T^{\mu}_{\nu}=0$, as can also explicitly checked taking the covariant derivative of the right-hand side
of \eq{coveq} and using the equations of motion for the auxiliary fields.

\section{Cosmology with the non-local Weyl-square term}\label{cosmology}

\subsection{Background evolution}\label{sect:back}

At the level of the background, the evolution is the same as in the model $\mu_2=0$, already studied in
\cite{Maggiore:2014sia}. For later use, let us recall the main results. We consider a flat FRW metric, and we use conformal time, 
$ds^2=a^2(\eta)(-d\eta^2+d\vx^2)$.
Then, the Einstein equations and the  evolution of the background configurations of the auxiliary fields $U$ and $S$, eq.~(\ref{eq1}),  can be recast in the form
\bees\label{h}
&&h^2=\frac{\Omega_M e^{-3 x}+\Omega_R e^{-4 x}+\left(\gamma_1/4\right) \bar{U}^2}{1+\gamma_1\left(-3\bar{V}'-3\bar{V}+\frac{1}{2}\bar{V}'\bar{U}'\right)}\,, \\
&&\bar{U}''+\left(3+\zeta\right)\bar{U}'=6\left(2+\zeta\right)\,, \label{barU}\\
&&\bar{V}''+\left(3+\zeta\right)\bar{V}'=\bar{U} h^{-2}\,,\label{barV}
\ees
where  the bar over $U$ and $V$ denotes their background value, the  prime is the derivatives with respect to $x=\log a$, and  $h={\cal H}/{\cal H}_0=H/H_0$, where $H=(1/a) da/dt$. As usual,
$\Omega_R$ and $\Omega_M$ are the energy fractions of radiation and matter at present time normalized with respect to $\rho_0=3H_0^2/\left(8\pi G\right)$. We also introduced the notation $\bar{V}\equiv H_0^2 \bar{S}$, as well as
$\zeta\equiv h'/h$ and
\be\label{defgamma1mu1}
\gamma_1\equiv \frac{2\mu_1}{3H_0^2}\, .
\ee
Eqs. (\ref{h})-(\ref{barV}) is a closed system and can be solved numerically given the initial conditions on the auxiliary fields $\bar{U}$, $\bar{V}$ fields and their first derivative. In particular, we assume  
$\bar{U}=\bar{V}=\bar{U}'=\bar{V}'=0$, at an initial time  deep in RD.
It is also useful to define an effective dark energy density $\rho_{\rm DE}$, rewriting the modified Friedman equation (\ref{h}) as
\be
h^2(x)=\Omega_M e^{-3 x}+\Omega_R e^{-4 x}+\frac{\rho_{\rm DE}}{\rho_0}\, ,
\ee
Similarly, we can define an effective DE pressure  $p_{\rm DE}$ from the trace of the $(ij)$  modified  Einstein equations, and the corresponding equation-of-state (EOS) parameter
$w_{\rm DE}\equiv p_{\rm DE}/\rho_{\rm DE}$. 

In {\LCDM} one  typically uses,  as  independent cosmological parameters,
the baryon density today $\omega_b=\Omega_bh_0^2$,  the cold dark matter density
$\omega_c=\Omega_{\rm c}h_0^2$, the Hubble parameter today 
$H_0=h_0\times 100\,{\rm km}\,{\rm s}^{-1}{\rm Mpc}^{-1}$, the amplitude of scalar perturbation $A_s$, the scalar spectrum index $n_s$ and the redshift at which the Universe is half-reionized $z_{\rm re}$. 
The dark energy density $\ola$ is then a derived parameter, fixed by the flatness condition. 
In the non-local model with $\mu_2=0$ the situation is similar. We can use the same set of independent cosmological parameter and
then $\mu_1$, or equivalently $\gamma_1$, is a derived parameter, fixed again from the condition  $\Omega_{\rm tot}=1$. 
The corresponding comparison with data for the $R\Box^{-2}R$ model has been performed in \cite{Dirian:2014bma},
implementing the cosmological perturbations of the non-local models into a Boltzmann code and performing parameter estimation. The  best-fit values for $\omega_c$, $\omega_b$ and $h_0$, using Planck 2013 CMB data, JLA SNe and BAO are $\omega_c=0.1204(14)$,  $\omega_b=2.197(25)\times 10^{-2}$ and $h_0=0.709(7)$, corresponding to
$\oma =(\omega_c+\omega_b)/h_0^2=0.283(9)$. The value of  $\gamma_1$ is then fixed so to reproduce this best-fit value,
which gives\footnote{In principle the value of $\gamma_1$ is determined by trial and errors so that the evolution produces the required value of $\ode$ and therefore of $\oma$ at the present time $x=0$. However, in the 
region $\oma\in [0.20,0.35]$, the required value of $\gamma_1$ is accurately reproduced by the fit
$\gamma_1 =0.0103959 + 0.00687851 \oma - 0.0598026 \oma^2 + 0.094128 \oma^3 - 
 0.0624636 \oma^4$ \cite{Dirian:2014ara}. Observe also that this value of $\gamma$ has been obtained using the Planck 2013 data. An updated analysis using Planck 2015 data will be presented in \cite{Diraninprep}.} 
$\gamma_1\simeq 9.286\times 10^{-3}$.

\subsection{Scalar-vector-tensor decomposition of the perturbations}\label{sect:deco}

We now turn to the study of the cosmological perturbations of this non-local model.  
We consider the following metric, in conformal time and longitudinal gauge
\be
ds^2=a^2(\eta)\left\{-\left(1+2\Psi\right)d\eta^2+2 w_i d\eta dx^i+\left[\left(1+2 \Phi\right) \delta_{ij}+2 h_{ij}\right]dx^i dx^j\right\}\,,
\ee
where $\Phi$ and $\Psi$ describe the scalar perturbations, the transverse vector $w_i$ describes vector perturbations, and the transverse-traceless tensor $h_{ij}$ describes tensor perturbations.
Let us recall that, for a generic anisotropic fluid, at first order in perturbation theory we have
$T^0_0=-(\bar{\rho}+\delta\rho)$, 
$T^0_i=(\bar{\rho}+\bar{p})v_i$ and 
$T^i_j=(\bar{p}+\delta p)\delta_j^i+\Sigma_j^i$,
where the density and the pressure have been perturbed around their background values, $\rho=\bar{\rho}+\delta{\rho}$ and 
$p\equiv \bar{p}+\delta p$. The pressure perturbation is proportional to the density perturbation,  $\delta p=c_s^2 \delta\rho$, where $c_s^2$ is the speed of sound of the fluid. We define as usual $\delta\equiv \delta\rho/\bar{\rho}$ and $\theta\equiv \delta^{ij}\partial_i v_j$. The anisotropic stress $\Sigma^i_j$ is a symmetric and traceless tensor, $\Sigma_i^i=0$. We consider the universe filled with radiation and non-relativistic matter, and therefore in the following we will set  $\Sigma_j^i= 0$.  
We expand the auxiliary fields $U$ and $S$ around some background configuration as
$U=\bar{U}+\delta U$ and $S=\bar{S}+\delta S$. In principle one can do the same with $U_{\mu\nu\alpha\beta}$ and $S_{\mu\nu\alpha\beta}$, parametrizing perturbations around some background configuration $\bar{U}_{\mu\nu\alpha\beta}$ and $\bar{S}_{\mu\nu\alpha\beta}$.   However, as we will explicitly show later, from the definitions (\ref{loc2}) it turns out that the background configurations $\bar{U}_{\mu\nu\alpha\beta}$ and $\bar{S}_{\mu\nu\alpha\beta}$ are vanishing on a cosmological FRW background. Therefore, the auxiliary tensor fields 
$U_{\mu\nu\alpha\beta}$ and $S_{\mu\nu\alpha\beta}$ do not enter the background equations and only affect  the  perturbations. 
This is consistent with what already observed: at the  background level the model (\ref{action}) is indistinguishable from the one with $\mu_2=0$. 
We also observe  that, as long as we are interested in  perturbations of first-order in the equations of motion,  the term $\mathcal{P}_{\mu\nu}$ 
in eq.~(\ref{coveq}) does not contributes, since it is quadratic in the tensor fields $U_{\mu\nu\alpha\beta}$ and $S_{\mu\nu\alpha\beta}$, which have vanishing background configurations on a FRW background for the metric. Computing the   Weyl tensor in terms of the metric perturbations we find, to first order
\be\label{C1}
C^0_{\,\,\,\,i0j}=-\frac{1}{2} D_{ij}\left(\Psi-\Phi\right)-\frac{1}{2}\pa_{\eta}w _{ij}+\frac{1}{2}\left(\partial_{\eta}^2+\n^2\right)h_{ij}\,,
\ee
\be
C^0_{\,\,\,\,ijk}=\left(\n_k w_{ij}-\n_j w_{ik}\right)+\frac{1}{4}\n^2\left(\delta_{ij}w_k-\delta_{ik} w_j\right)+\left(\n_j \pa_{\eta}h_{ik}-\n_k \pa_{\eta}h_{ij} \right)\,,
\ee
where
\be
D_{ij}\equiv\n_i\n_j-\frac{1}{3}\delta_{ij}\n^2\,,\hspace{2 em}w_{ij}\equiv\n_{(i}w_{j)}\,,
\ee
and, for any tensor $A_{\mu\nu}$,  we define $A_{(\mu\nu)}\equiv (1/2)(A_{\mu\nu}+A_{\nu\mu})$ and
$A_{[\mu\nu]}\equiv (1/2)(A_{\mu\nu}-A_{\nu\mu})$.
The other components of the Weyl tensor follow from the symmetry relations
$C_{\mu\nu\alpha\beta}=C_{[\mu\nu][\alpha\beta]}=C_{\alpha\beta\mu\nu}$ and $C_{\mu\{\nu\alpha\beta\}}=0$
(where, for any tensor $A_{\mu\nu\rho}$, $A_{\{\mu\nu\rho\}}\equiv A_{\mu\nu\rho}+A_{\nu\rho\mu}+A_{\rho\mu\nu}$),
which in 3+1 dimensions also imply 
\be\label{C3}
C_{ijkl}=\delta_{jl}\,C_{0i0k}+\delta_{ik}\,C_{0j0l}-\delta_{il}\,C_{0j0k}-\delta_{jk}\,C_{0i0l}\,.
\ee
Note that, since we are retaining only  terms of first-order in the metric perturbations, the components of the Weyl tensor  may be raised and lowered using the unperturbed metric. 
For the independent components of the $U_{\mu\nu\alpha\beta}$ tensor, i.e. $U_{0i0j}$ and $U_{0ijk}$,  eq.~(\ref{eq1}) on a FRW background is equivalent to the following set of coupled differential equations
\begin{align}\label{eqU0ijk}
\pa^2_{\eta}U_{0ijk}-6 \mathcal{H} \pa_{\eta}U_{0ijk}+&\left(2\mathcal{H}^2-4\pa_{\eta}\mathcal{H}\right)U_{0ijk}=\n^2 U_{0ijk}-a^2 C_{0ijk}+\nn\\
&+2\mathcal{H}\delta_{ij}\partial_{q}U_{0q0k}-2\mathcal{H}\delta_{ik}\partial_q U_{0q0j}-4\mathcal{H}\partial_j U_{0i0k}+4\mathcal{H}\partial_k U_{0i0j}\,.
\end{align}
\be\label{eqU0i0j}
\pa^2_{\eta}U_{0i0j}-6 \mathcal{H} \pa_{\eta}U_{0i0j}+\left(2\mathcal{H}^2-4\pa_{\eta}\mathcal{H}\right)U_{0i0j}=\n^2 U_{0i0j}+2\mathcal{H}\partial_{q}U_{0jiq}+2\mathcal{H}\partial_q U_{0ijq}-a^2 C_{0i0j}\,,
\ee
where, in the first equation, we used the fact that
the tensor $U_{ijkl}$ satisfies the same symmetry relation as the Weyl tensor, so in particular
also the analogous of the relation
(\ref{C3}). 
Analogously one has, for $S_{\mu\nu\rho\sigma}$,
\begin{align}\label{eqS0ijk}
\pa^2_{\eta}S_{0ijk}-6 \mathcal{H} \pa_{\eta}S_{0ijk}+&\left(2\mathcal{H}^2-4\pa_{\eta}\mathcal{H}\right)S_{0ijk}=\n^2 S_{0ijk}-a^2 U_{0ijk}+\nn\\
&+2\mathcal{H}\delta_{ij}\partial_{q}S_{0q0k}-2\mathcal{H}\delta_{ik}\partial_q S_{0q0j}-4\mathcal{H}\partial_j S_{0i0k}+4\mathcal{H}\partial_k S_{0i0j}\,,
\end{align}
\be\label{eqS0i0j}
\pa^2_{\eta}S_{0i0j}-6 \mathcal{H} \pa_{\eta}S_{0i0j}+\left(2\mathcal{H}^2-4\pa_{\eta}\mathcal{H}\right)S_{0i0j}=\n^2 S_{0i0j}+2\mathcal{H}\partial_{q}S_{0jiq}+2\mathcal{H}\partial_q S_{0ijq}-a^2 U_{0i0j}\,.
\ee
We observe that, as already mentioned in section \ref{cosmology}, if we solve eqs. (\ref{eqU0i0j})-(\ref{eqS0ijk}) for the background configurations of the auxiliary fields, the source term $C_{0ijk}$ vanishes and we get as a solution $\bar{U}_{\mu\nu\alpha\beta}=0$ and $\bar{S}_{\mu\nu\alpha\beta}=0$.

Note that all equations above involve the same differential operator in conformal time
\be
{\cal T}[X]\equiv \pa^2_{\eta}X-6 \mathcal{H} \pa_{\eta}X+(2\mathcal{H}^2-4 \pa_{\eta}\mathcal{H})X\,,
\ee
which may allow for unstable solutions.
This can be seen explicitly by using $x=\log a$ as time variable. Then  the differential operator becomes
\be\label{N1}
{\cal T}[X]\rightarrow\hat{\cal T}[X]\equiv \mathcal{H}^2\left[X''+\left(\zeta-5\right) X'-2\left(2\zeta+1\right)X\right]\,.
\ee
As in section~\ref{sect:Rmnterms}, we can take $\zeta(x)\simeq \zeta_0$ as
approximatively constant in the various cosmological epochs, with  $\zeta_0=\{-2\,,-3/2\,,0\}$ in radiation, matter and de~Sitter, respectively. 
One then finds that in RD there is a mode growing as $e^{6x}=a^6$ which might suggest that the model is badly unstable; however, we will see that the auxiliary fields suffering this instability enter in the Einstein equations with a suppression factor which is just $a^{-6}$, so an explicit inspection of the Einstein equations is necessary in order to understand the behavior of the system.

We now parametrize the various components of  $U_{\mu\nu\alpha\beta}$ and $S_{\mu\nu\alpha\beta}$ in terms of scalar, vector and tensor degrees of freedom. The five degrees of freedom characterizing the symmetric and traceless tensors $U_{0i0j}$ and $S_{0i0j}$ can be decomposed in the usual way into a scalar, a transverse vector, and a transverse traceless symetric tensor:
\be\label{U1}
U_{0i0j}= D_{ij} u+\n_{(i}u_{j)}+u_{ij}\,,
\ee
\be\label{S1}
S_{0i0j}= D_{ij} s+\n_{(i}s_{j)}+s_{ij}\,,
\ee
with $\n_i u^i=\n_i s^i=u_i^i=s_i^i=\n_i u^{ij}=\n_i s^{ij}=0$.  
For the other components, one may notice that $S_{0ijk}$  enters in the Einstein equations only through the combinations
$S_{0(ij)k,k}\equiv  S_{ij}$, and therefore $U_{0ijk}$ only enters through the combination 
$U_{0(ij)k,k}\equiv  U_{ij}$. These combinations are
symmetric, traceless and with one vanishing longitudinal component ($S_{ij,ij}=0$), so they  only carry four degrees of freedom (the scalar is missing). Then, they   are naturally parametrized as
\be\label{S1}
S_{ij}= \n_{(i}\sigma_{j)}+\sigma_{ij}\,.
\ee
Observe that $\n_i\n_jS_{ij}=0$ to first order in perturbation theory since, on FRW, the Christoffel symbol $\Gamma^k_{ij}=0$. Similarly, we write
\be\label{U1}
U_{ij}= \n_{(i}v_{j)}+v_{ij}\,,
\ee
with $\n_i v^i=\n_i \sigma^i=v_i^i=\sigma_i^i=\n_i v^{ij}=\n_i \sigma^{ij}=0$.  
We  now have all the elements for writing down the Einstein equations to first order in the cosmological perturbations, and decomposing them  
in the usual way into scalar, vector and tensor components. 
We also go in Fourier space, denoting the comoving momentum by $k$. We introduce $\hat{k}\equiv k/(aH)$, and we 
use $x=\log{a}$ as time variable, with $df/dx=f'$. It is also useful to introduce the  dimensionless couplings 
\be
\gamma_1\equiv \frac{2 \mu_1}{3H_0^{2}}\, ,\qquad
\gamma_2\equiv \frac{2 \mu_2}{3H_0^{2}}\, ,
\ee
 and to rescale the fields as follows,
\be\label{fieldef}
V=H_0^2 S\,,\hspace{1 em}\hat{s}=H_0^4 s\,,\hspace{1 em}\hat{u}=H_0^2 u\,,\hspace{1 em}\hat{s}_{ij}\equiv H_0^2 s_{ij}\,,\hspace{1 em}\hat{\sigma}_{ij}\equiv H_0 \sigma_{ij}\,,\hspace{1 em}\hat{v}_{ij}\equiv \frac{v_{ij}}{H_0}\,,
\ee

\subsection{Scalar sector}\label{sect:scal}

In the scalar sector, we can take as independent equations the trace and the traceless parts of the $(ij)$ component of the perturbed Einstein equations. The former gives
\begin{align}\label{eqs1}
&\left(1-3 \gamma_1 \bar{V}\right)\left[\Phi''+\left(\zeta+3\right)\Phi'-\Psi\left(3+2\zeta\right)
-\Psi'+\frac{\hat{k}^2}{3}\left(\Psi+\Phi\right)\right]\nn\\
&-\frac{3}{2} \gamma_1\Bigg[ \frac{1}{2 h^2}\bar{U}\delta U-2\Psi \bar{V}''
+\[  2\Phi'-2(\zeta+2)\Psi-\Psi'-\Psi \bar{U}'  \]\bar{V}'+{\delta V}''
\nn\\
&\hspace*{12mm}+\left(\zeta+2\right){\delta V}'+\frac{2\hat{k}^2}{3}\delta V+\left(3+2\zeta\right)\delta V+\frac{1}{2}\left(\bar{U}'{\delta V}'+\bar{V}'{\delta U}'\right)\Bigg]\nn\\
&+\frac{2}{3}\frac{\gamma_2}{a^2}h^2\hat{k}^4 s=-4\pi G \frac{\delta p}{H^2}\,,
\end{align}
and the latter
\be\label{eqs2}
\left(1-3\gamma_1\bar{V}\right)\Psi=-\left(1-3\gamma_1\bar{V}\right)\Phi+3\,\gamma_1\, \delta V-\frac{6\,\gamma_2}{a^2}h^2\,\left[{\hat{s}}''+\left(\zeta-3\right){\hat{s}}'-\left(\frac{5}{4}\zeta-2\right)\hat{s}+\frac{\hat{k}^2}{3}\hat{s}\right]\,.
\ee
As usual, the $(00)$ component of the Einstein equation is not independent from  these equations.
The equations for the auxiliary fields can be written as
\begin{align}\label{eqs3}
&{\delta U}''+\left(3+\zeta\right){\delta U}'+\hat{k}^2\delta U-2\Psi \bar{U}''-\left[2\left(3+\zeta\right)\Psi+\Psi'-3\Phi'\right]\bar{U}'=\nn\\
&=2\hat{k}^2\left(\Psi+2\Phi\right)+6\left[\Phi''+\left(\zeta+4\right)\Phi'\right]-6\left[\Psi'+2\left(\zeta+2\right)\Psi\right]\,,
\end{align}
\be\label{eqs4}
{\delta V}''+\left(\zeta+3\right){\delta V}'+\hat{k}^2 \delta V -2\Psi \bar{V}''-\left[2\left(\zeta+3\right)\Psi+\Psi'-3\Phi'\right]\bar{V}'=\frac{\delta U}{h^2}\,,
\ee
\be\label{eqs5}
{\hat{s}}''+\left(\zeta-5\right){\hat{s}}'-2\left(2\zeta+1\right)\hat{s}+\hat{k}^2 s=-\frac{1}{h^2}\hat{u}\,,
\ee
\be\label{eqs6}
{\hat{u}}''+\left(\zeta-5\right){\hat{u}}'-2\left(2\zeta+1\right)\hat{u}+\hat{k}^2 u=-\frac{a^2}{2}\frac{1}{h^2}\left(\Psi-\Phi\right)\,.
\ee
From the covariant conservation of the energy-momentum tensor, $\n^{\mu}T_{\mu\nu}=0$, we find
\bees
&&\delta'=-\left(3\Phi'+\hat{\theta}\right)\left(1+w\right)-3\delta \left(c_s^2-w\right)\,,\\
&&\hat{\theta}' =-\left(2-3w+\zeta+\frac{w' }{1+w}\right)\hat{\theta}+\hat{k}^2\left(\Psi+\sigma+\frac{c_s^2}{1+w}\,\delta\right)\,,
\ees
where $w$ is defined by $\bar{p}=w\bar{\rho}$, while $c_s^2$ from $\d p=c_s^2\d\rho$.\footnote{Recall that, for a barotropic fluid, $c_s^2=w+(\bar{\rho}/\bar{\rho}' )  w'$.  The term $w'$ is only important at the transition between different phases. In a given phase, such as RD or MD, when $w$ is constant to good accuracy,
$c_s^2\simeq w$.}
These equations are independent of the specific dark energy content, since they simply express the conservation of $T_{\mu\nu}$. For matter-radiation  fluids with no energy exchange among them, we have as usual
\be\label{eqs7}
\delta'_M=-(3\Phi'+\hat{\theta}_M)\,,
\ee
\be\label{eqs8}
\hat{\theta}' _M=-\left(2+\zeta\right)\hat{\theta}_M+\hat{k}^2 \Psi\,,
\ee
\be\label{eqs9}
\delta'_R=-\frac{4}{3}(3\Phi'+\hat{\theta}_R)\,,
\ee
\be\label{eqs10}
\hat{\theta}' _R=-\left(1+\zeta\right)\hat{\theta}_R+\hat{k}^2\left(\Psi+\frac{\delta_R}{4}\right)\,.
\ee
We set to zero the amplitude of the auxiliary fields and of their derivatives at an initial time deep into RD. For the matter and metric variables, we take the standard adiabatic initial conditions expanded to second order (see e.g. \cite{Ruth_book}, or eqs.~(6.1)-(6.3) of \cite{Dirian:2014ara}).
For the spectral index $n_s$ and for the amplitude of the gravitational potential $\delta_H$, we take the  values, $n_s\simeq 0.96$\,, $\delta_H^2\simeq 3.2\cdot 10^{-10}$. We can now solve the set of coupled differential equations (\ref{eqs1})-(\ref{eqs6}),  (\ref{eqs7})-(\ref{eqs10}),  for different values of $k$.  

Let us recall that the modes relevant for the comparison with CMB and with structure formation data correspond roughly to the range $k=(10^{-3}-10)\, h/{\rm Mpc}$, and that in {\LCDM} modes with $k\,\lsim\,  0.3\, h/{\rm Mpc}$ remain in the linear regime up to the present epoch, while modes with  $k\,\gsim\,  0.3\, h/{\rm Mpc}$ become non-linear. In terms of the quantity $k/H_0$, the range relevant for CMB and structure formation is roughly $k/H_0\in [3,3\times 10^4]$, and non-linear scales correspond to $k/H_0\,\gsim\,  900$. 

\begin{figure}[t]
\centering
\includegraphics[width=0.5\columnwidth]{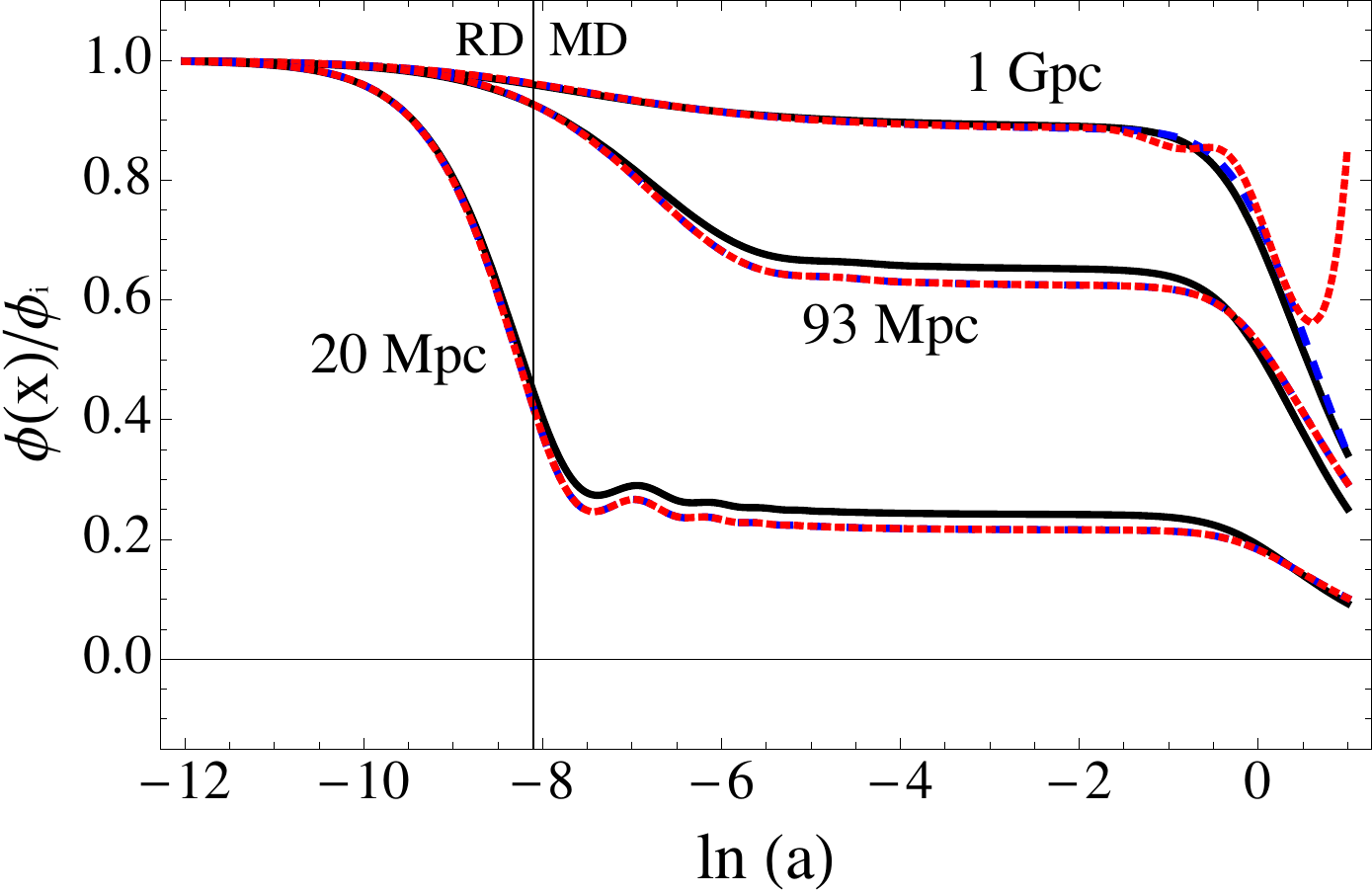}
\caption{\label{fig:scalars} Scalar perturbations for three different comoving reduced wavelength $\lbar=1/k$. For each wavelength we plot the result for {\LCDM} (black solid line), the pure $R\Box^{-2}R$ model with $\mu_1=0.014 H_0^2$ and
$\mu_2=0$ (blue, dashed), and the model with $\mu_1=\mu_2=0.014 H_0^2$ (red, dotted). For 
$\lbar=20$~Mpc and for $\lbar=93$~Mpc the blue and the red curves are basically indistinguishable on this scale, while for $\lbar =1$~Gpc the blue and black lines are almost  indistinguishable.}
\end{figure}

Our numerical results are shown in Fig.~\ref{fig:scalars}. In this figure we show the results for  three different values of the (reduced) comoving wavelength $\lbar=1/k=\{20, 93,1000\}\, {\rm Mpc}$. These values are representative of different kind of behaviors. Indeed, 
the mode with $\lbar =20$~Mpc (i.e $k/H_0\simeq 214$, or $k\simeq 0.071 \, {\rm Mpc}/h_0$) re-enters the horizon deep in RD.  
The mode with 
$\lbar =93$~Mpc (i.e $k/H_0\simeq 46$, or $k\simeq 0.015\,  {\rm Mpc}/h_0$) is the one that  (when $h_0=0.7$ and $\oma=0.3$) re-enters the horizon exactly at the RD-MD transition.
Finally, the mode with $\lbar=1$~Gpc (i.e $k/H_0\simeq 4.3$, or $k\simeq 0.001\,  {\rm Mpc}/h_0$) re-enters the horizon only near the present epoch.

For each of these wavelengths  we show the standard {\LCDM} result (black solid line), the result for the pure $R\Box^{-2}R$ model (blue, dashed line)  with  $\mu_1=0.014 H_0^2$, which corresponds to the  best fit value for $\gamma$ given in Sect.~\ref{sect:back},  and the result for the model with also the Weyl-square term, where we set  for definiteness $\mu_1=\mu_2=0.014 H_0^2$ 
(red, dotted).
We see that, compared to the pure $R\Box^{-2}R$ model, and up to the present time, only the largest modes are affected by the addition of the Weyl-square term. 
Even for these very long wavelengths, which are  the largest observable with CMB anisotropies, the difference with  the $R\Box^{-2}R$ model 
 is negligible during MD, and  only shows up as a difference of at most (1-2)\% percent near the recent epoch, $x=0$ (although scalar perturbations in the Weyl model  will start to show an instability in the cosmological future, $x>0$). 
 
This means that, compared to the  pure $R\Box^{-2}R$ model, the addition of the Weyl-square term has basically no effect on SNe, BAO, and structure formation data while, in the CMB, it only affects the late integrated Sachs-Wolfe (ISW) effect, just at the level of order (1-2)\% percent. The late ISW effect only contributes to the very low multipoles, and even for these multipoles it is sub-leading with respect to the  (non-integrated) Sachs-Wolf effect. Thus, among all cosmological observations testing scalar perturbations, the inclusion of the Weyl-square term will only affects the lowest CMB multipoles, and even for these  it  will only result in an overall  deviation of the $C_l$ at the sub-percent level.  Therefore,  in the presence of the non-local Weyl-square term, scalar perturbations remain are well-behaved (at least up to the present time) and in principle consistent with the data.

Conceptually, it is however interesting to observe that, as we extend the time integration more into the future, we start to see more and more deviations in the scalar perturbations of the Weyl-square model, compared to {\LCDM} or to the pure $R\Box^{-2}R$ model. The behavior of the long wavelength modes can be understood  analytically, and we perform the corresponding analysis in App.~\ref{sect:appSH}.

\subsection{Tensor sector}\label{sect:tenssect}

We next turn to tensor perturbations, where we will see that the inclusion of the non-local Weyl-square term can be more problematic.
In the tensor sector we have to solve 5 coupled second-order differential equations in the variables $h_{ij}$, $u_{ij}$, $v_{ij}$, $\sigma_{ij}$, $s_{ij}$, coming from the TT part of the Einstein equations and of the equations for the auxiliary fields. With the field redefinition (\ref{fieldef}), the relevant equations are
\be\label{stens}
{\hat{s}}''_{ij}+\left(\zeta-5\right){\hat{s}}'_{ij}-\left(2+4\zeta\right)s_{ij}+\hat{k}^2\hat{s}_{ij}=\frac{4}{a h}\hat{\sigma}_{ij}-\frac{1}{h^2}u_{ij}\,,
\ee
\be\label{utens}
{\hat{u}}''_{ij}+\left(\zeta-5\right){\hat{u}}'_{ij}-\left(2+4\zeta\right)u_{ij}+\hat{k}^2 u_{ij}=\frac{4}{a h}\hat{v}_{ij}+\frac{a^4}{2}\left(h''_{ij} +\left(1+\zeta\right)h'_{ij}-\hat{k}^2 h_{ij}\right)\,,
\ee
\be\label{sigmatens}
\hat{\sigma}''_{ij}+\left(\zeta-5\right)\hat{\sigma}'_{ij}-\left(2+4\zeta\right)\sigma_{ij}+\hat{k}^2 \hat{\sigma}_{ij}=-4 a\, h \,\hat{k}^2\, \hat{s}_{ij}-\frac{1}{h^2}\hat{v}_{ij}\,,
\ee
\be\label{vtens}
\hat{v}''_{ij}+\left(\zeta-5\right)\hat{v}'_{ij}-\left(2+4\zeta\right)v_{ij}+\hat{k}^2 \hat{v}_{ij}=-4 a\, h \,\hat{k}^2 \hat{u}_{ij}+a^5 \hat{k}^2\, h\, h'_{ij} \,.
\ee
\begin{align}\label{htens}
&\left(1-3\gamma_1 \bar{V}\right)\left(h''_{ij} +\left(\zeta+3\right)h'_{ij} +\hat{k}^2 h_{ij}\right)-3\gamma_1 \bar{V}'h'_{ij} -\nn\\
&-\frac{6 \gamma_2}{a^4}\left[{\hat{s}}''_{ij}+\left(\zeta-3\right) {\hat{s}}'_{ij}+2\left(1-\zeta\right)\hat{s}_{ij}-\hat{k}^2 \hat{s}_{ij}+\frac{2}{a h}\hat{\sigma}'_{ij}-\frac{4}{a h} \hat{\sigma}_{ij}\right]=0\,,
\end{align}
where in the last equation we have neglected the contribution coming from the anisotropic stress. 

As usual, we separate the Fourier components $h_{ij}(\eta,\vk)$ into its plus and cross polarizations,  
 \be
h_{ij}(\eta,\vk) =  h_+(\eta,\vk)e^{+}_{ij}(\vk)  + h_{\times}(\eta,\vk)e^{\times}_{ij}(\vk)\, ,
 \ee
and similarly for all other transverse-traceless tensors $\hat{s}_{ij},  \hat{u}_{ij}$, etc. The equations for the two polarizations are the same, and we denote generically their amplitudes as $h_g$ (not to be confused with the reduced Hubble parameter $h=H/H_0$), $s_g$, $u_g$, etc.

\begin{figure}[t]
\centering
{\includegraphics[scale=0.45]{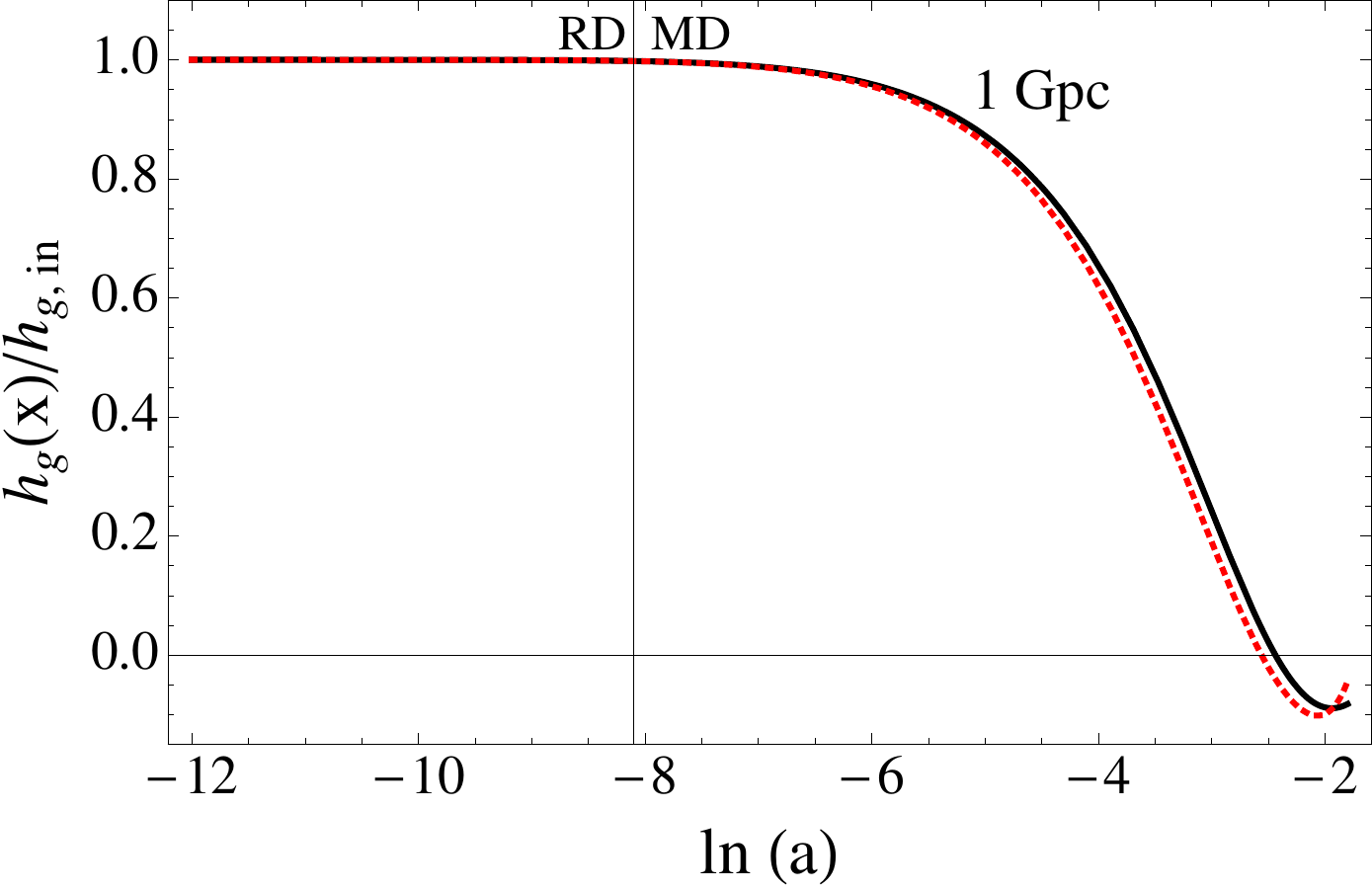}}\qquad
{\includegraphics[scale=0.45]{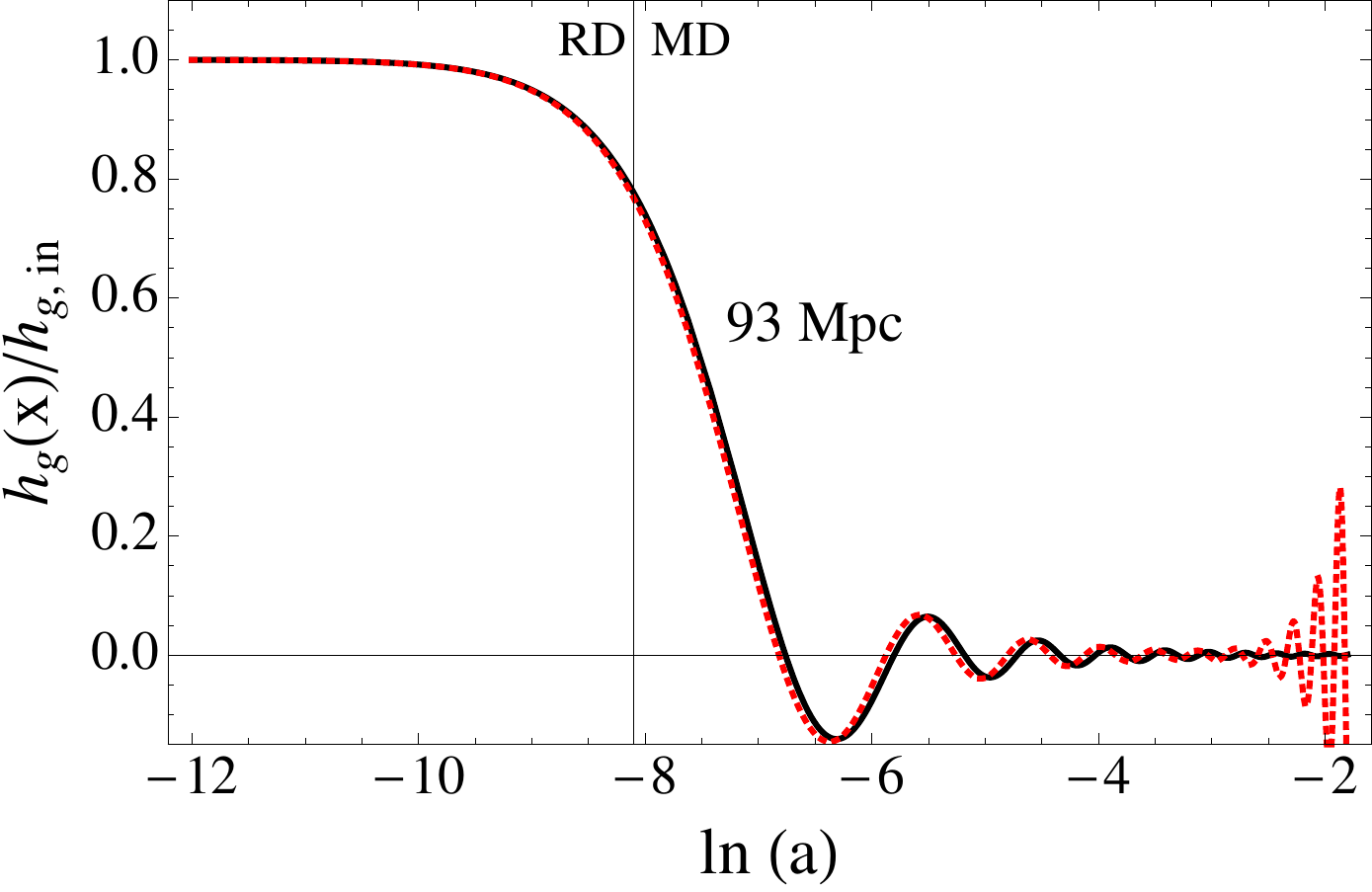}}
{\includegraphics[scale=0.45]{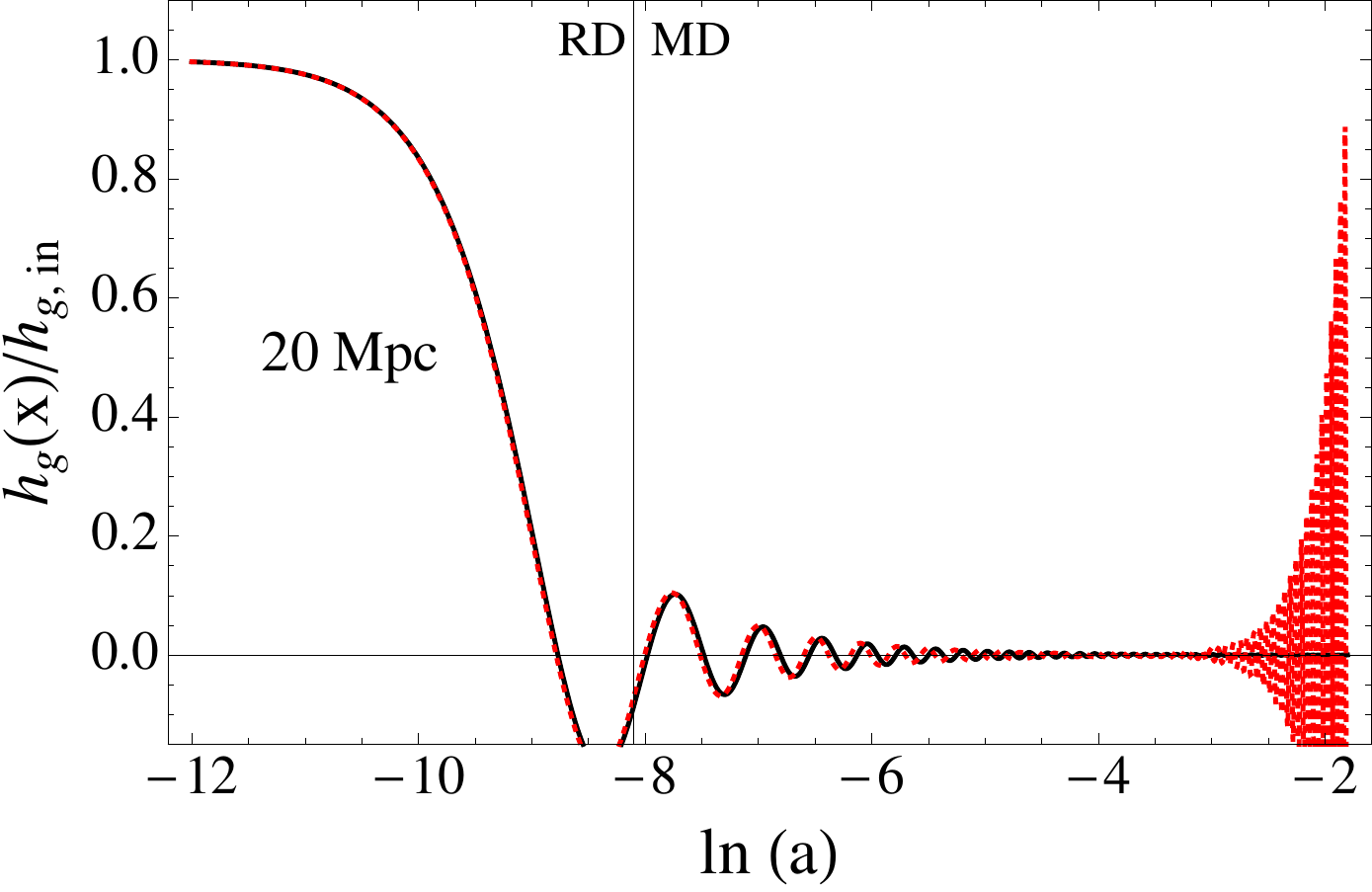}}
\caption{\label{fig:tensor} \small The evolution of tensor perturbations, 
for   three different comoving reduced wavelength. For each wavelength we plot the result for {\LCDM} (black solid line) and for the model with $\mu_1=\mu_2=0.014 H_0^2$ (red, dotted). 
}
 \end{figure}   

\begin{figure}[t]
\centering
{\includegraphics[scale=0.45]{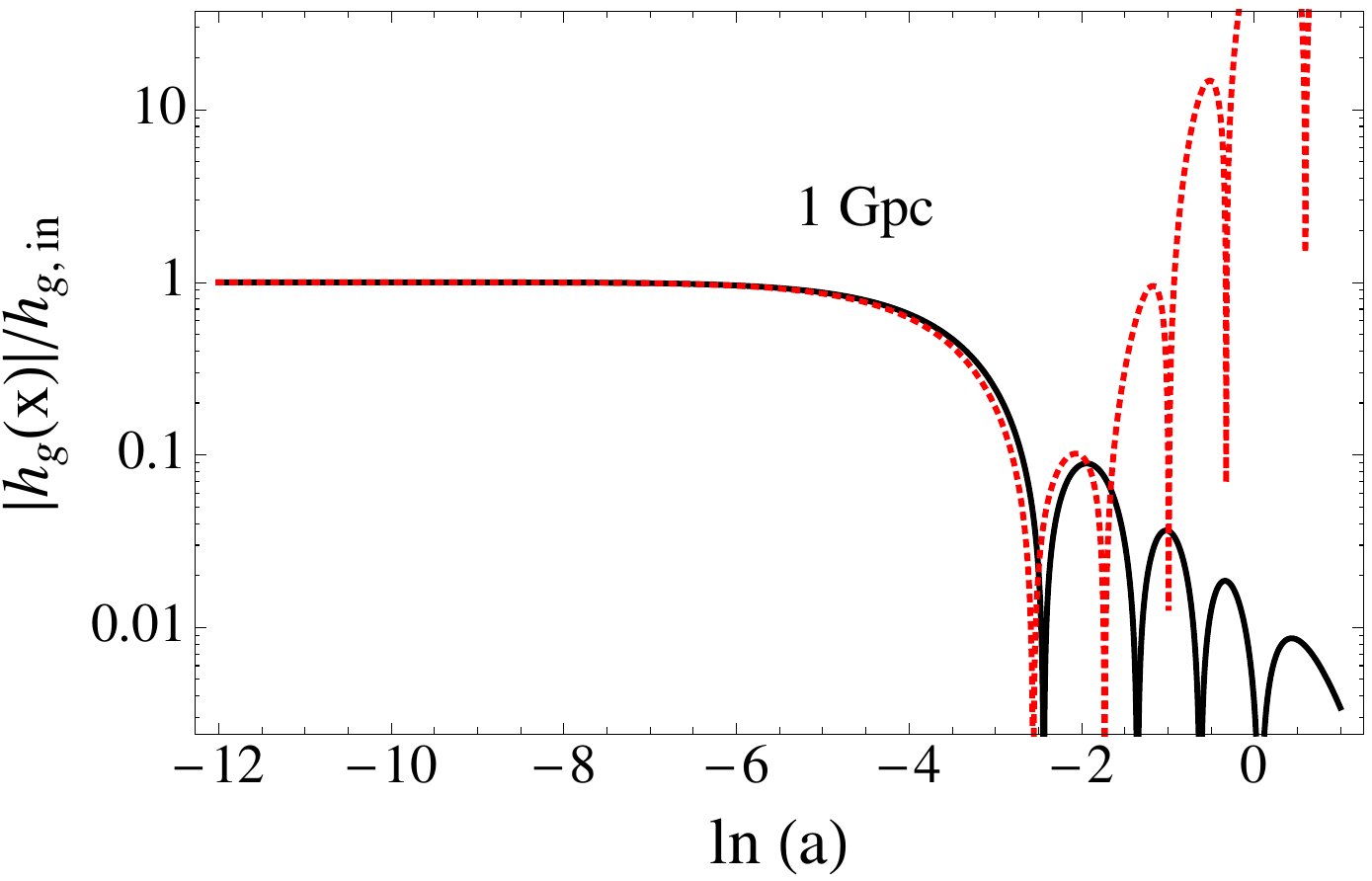}}\qquad
{\includegraphics[scale=0.45]{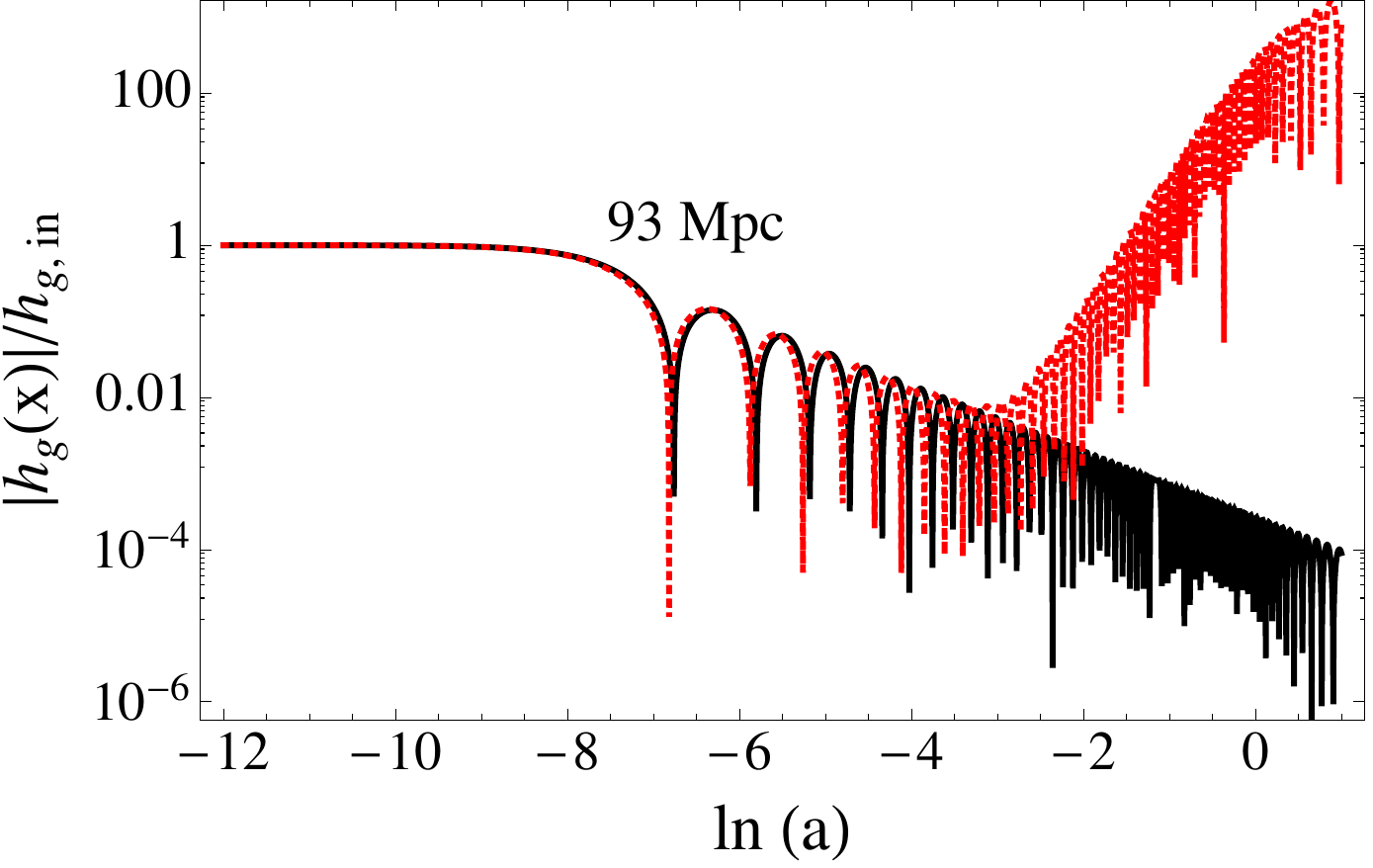}}
{\includegraphics[scale=0.45]{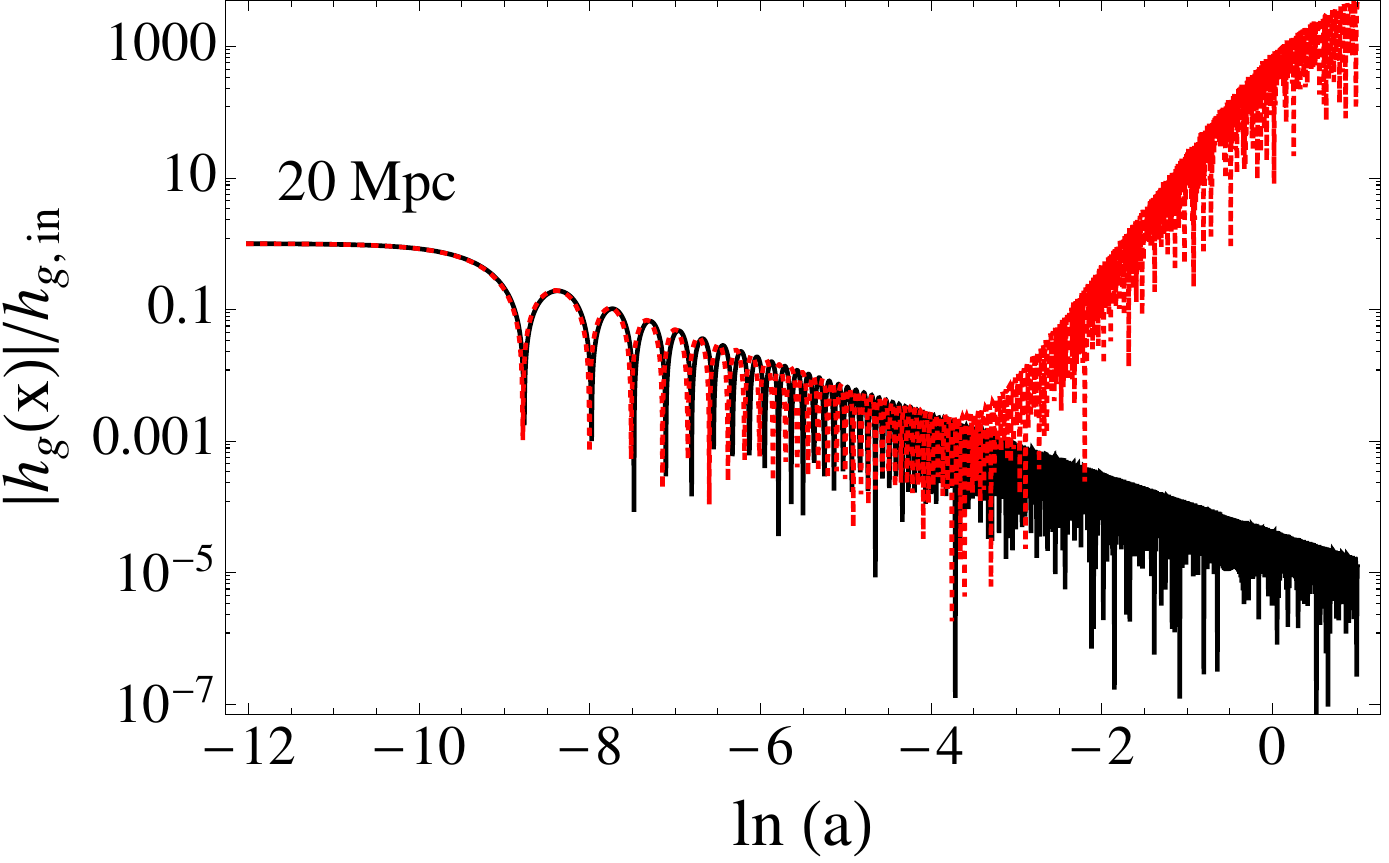}}
\caption{\label{fig:tensorLog} \small As in Fig.~\ref{fig:tensor}, on a logarithmic vertical  scale, and extending the horizontal axis to $\log a=1$. Observe the different vertical scale among the figures.
}
 \end{figure}

We solve numerically \eqst{stens}{htens} starting from an initial redshift $z_{i}=10^{9}$ in radiation domination, with GR-like initial conditions for the metric tensor mode $h_g$, i.e. $h_g(z_{i})=1$ and 
 $h'_g(z_{i})=0$,  and vanishing initial conditions for the auxiliary fields $u_g$, $v_g$, $s_g$ and $\sigma_g$ and their first derivatives. 
The result of the numerical integration   
is shown  in Fig.~\ref{fig:tensor} where, for clarity, we show separately the result for
the same three comoving wavelengths already used
in Fig.~\ref{fig:scalars} for scalar perturbations. For each wavelength we now display only the standard {\LCDM} result (black solid line) and the result for the Weyl-square model with $\mu_1=\mu_2=0.014 H_0^2$ (red, dotted), since the result for the pure $R\Box^{-2}R$ model, on this scale, are hardly distinguishable from the result for {\LCDM}. We now find the surprise that, in  the Weyl-square model, tensor modes become unstable. The instability starts some time after the RD-MD transition and, as we will show analytically in App.~\ref{App:tensor}, is due to the fact that the auxiliary fields, which are zero during RD, gradually become non-vanishing in MD, and when they are sufficiently large they begin to source the instability. We also observe that the instability is stronger for the short wavelengths, i.e. large frequencies. This is due to the fact that it is actually the time derivative of the auxiliary fields that sources the instability, so the effect is larger for the modes oscillating with large frequency. Observe also that in Fig.~\ref{fig:tensor} we only plot the result up to $\log a=-1.8$. If we extended these plots up to the present time $\log a=0$, the instability would bring the amplitude well outside the vertical scale of the plots. The corresponding plots on a logarithmic vertical scale, extended up to the $\log a=1$ in the cosmological future, are shown in 
Fig.~\ref{fig:tensorLog}.

To assess whether this instability in the tensor sector rules out the model which features the non-local Weyl square term, we need to make assumptions on the initial spectrum of perturbation, $h_{g, \rm in}(k)$, whose subsequent evolution is then amplified as  in Fig.~\ref{fig:tensorLog}. This initial spectrum depends on the theory of the primordial Universe. Some alternatives to the standard slow-roll inflationary paradigm predict a negligible primordial tensor spectrum. For instance, 
the pre-big-bang model \cite{Gasperini:2002bn} predicts a primordial relic GW spectrum that,  in some range of its parameter space, can be sizable, but for other values of the parameters can be totally negligible
\cite{Brustein:1995ah,Buonanno:1996xc}. As another example, the  ekpyrotic model  predicts a negligible amount of primordial tensor perturbations~\cite{Khoury:2001wf,Boyle:2003km}. However, to generate an adiabatic spectrum of scalar perturbations,    the pre-big-bang must appeal to mechanisms such as the decay of massive axions, leading to the `curvaton' mechanism~\cite{Enqvist:2001zp} which, even if potentially viable, are less natural than the generation of adiabatic perturbations in standard slow-roll inflation.

If we stick to the standard inflationary paradigm, the amplitude of the primordial tensor perturbations is characterized by a tensor-to-scalar ratio $r$, with $r<O(0.1)$ to comply with the joint BICEP2/Keck Array-Planck analysis \cite{Ade:2015tva}. Several single-field inflationary models give predictions close to this experimental bound (or in excess of it, in which case they are ruled out). In any case, even the inflationary models that give a small primordial GW spectrum, 
such as the Starobinsky model, still predict values of $r$ at least of order $10^{-3}$.

In terms of the GW energy density per unit logarithmic interval of frequency, defined as usual as 
$\hogw (f)=(1/\rho_c)d\rho_{\rm GW}/d\log f$ (where the frequency $f$ today is related to $k$ by $k=2\pi f$),
the standard functional dependence of the inflationary prediction is $\hogw(f)\propto f^{-2}$ for frequencies $f$ in the range $3\times 10^{-18}\, {\rm Hz}<f<f_{\rm eq}$, where the condition $f>3\times 10^{-18}$ imposes that the GW is inside the horizon today, while $f_{\rm eq} \simeq 1.6\times 10^{-17}\, {\rm Hz}$ is the value of the frequency of the mode that re-enters the horizon at the RD-MD transitions. Modes with $f>f_{\rm eq}$ re-entered the horizon already during RD, and for these frequencies $\hogw (f)\propto f^{n_T}$, where $n_T=-r/8$, so in this regime the spectrum is basically flat (see e.g. \cite{Maggiore:1999vm} for review).

In the presence of the non-local Weyl-square term, this GW spectrum is further amplified by its evolution during MD, as shown in Fig.~\ref{fig:tensorLog}. We can  compute this extra amplification by comparing mode by mode the {\LCDM} result with  the result in the presence of the non-local Weyl-square term. We find that, because of this further amplification, in the regime $3\times 10^{-18}\, {\rm Hz}<f<f_{\rm eq}$ now $\hogw (f)$  grows as $f^3$, while in the regime $f>f_{\rm eq}$ it grows as $f^4$. 

\begin{figure}[t]
\centering
{\includegraphics[scale=0.9]{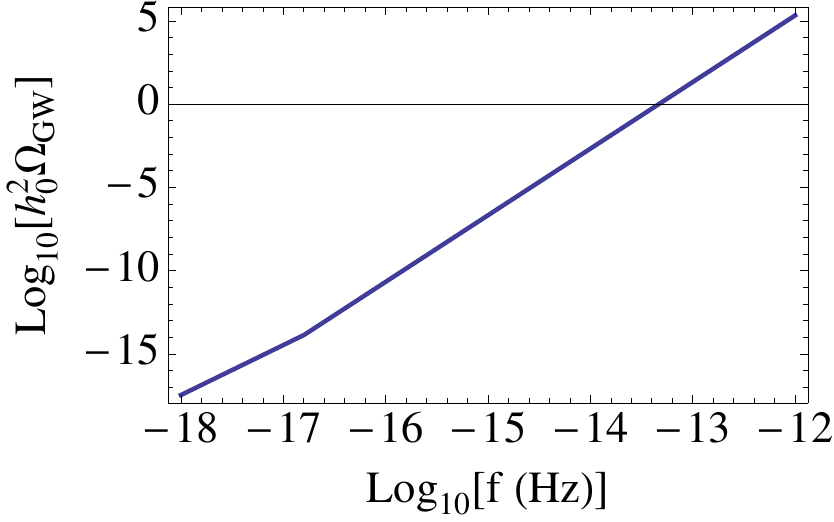}}
\caption{\label{fig:Omega_vs_f} \small The value of $\hogw (f)$ for a primordial inflationary spectrum further amplificated during the evolution to the present time by the non-local Weyl square term.
}
 \end{figure}   

Let us denote by  $f_*=k_*/(2\pi)$ the value of the frequency corresponding to the  pivot scale used by Planck. For the typical value $k_*=0.002\, {\rm Mpc}^{-1}$ we have 
$f_*\simeq 3.09\times 10^{-18}\, {\rm Hz}$. We denote by $\Omega_*$ the corresponding value of 
$\Omega_{\rm GW}(f=f_*)$. Then, the form of the inflationary spectrum today, after the further amplification by the Weyl-square term, is
\be
h_0^2\Omega_{\rm GW}(f)=h_0^2\Omega_*\times \left\{
\begin{array}{lc}
\(\frac{f}{f_*}\)^3&\hspace{10mm} 3\times 10^{-18}\, {\rm Hz}<f<f_{\rm eq}\, ,\\
\\
\(\frac{f_{\rm eq}}{f_*}\)^3\,  \(\frac{f}{f_{\rm eq}}\)^4 &\hspace{10mm} f>f_{\rm eq}\, .
\end{array}
\right.
\ee
Even assuming a very low value of $h_0^2\Omega_*$, such as $h_0^2\Omega_*=10^{-16}$ (which
corresponds to a tensor-to-scalar ratio $r\sim 10^{-4}$), 
such an extremely blue spectrum wildly violates all existing bounds on $\hogw (f)$, coming e.g. from pulsar timing, extra radiation or big-bang nucleosynthesis, and even quickly becomes larger than one, as we see  from Fig.~\ref{fig:Omega_vs_f}. In principle such a behavior will continue until the intrinsic cutoff of the inflationary spectrum is met, e.g. up to $f=O(10^9)$~Hz for typical GUT scale inflation.\footnote{Actually, our numerical integration only works up to momenta $k\simeq 10^4 H_0$, i.e. frequencies of order $f\simeq 3.5\times 10^{-15}$~Hz. For larger frequencies it becomes numerically difficult to follow reliably the growth of the tensor perturbations. However, such frequencies are already well in the range of frequencies that entered the horizon during RD, so we do not expect a further change of regime until the inflationary cutoff is met.}
In this figure we used the result obtained for $\mu_2=\mu_1=0.014 H_0^2$ and it is worth reminding that, contrarily to $\mu_1$ which is fixed by the background evolution, $\mu_2$ is in principle a free parameter. However it is clear that, unless $\mu_2$ is tuned to ridiculously small values, the resulting GW spectrum is unacceptably large.

The conclusion from this analysis is that, in a cosmological model in which the primordial tensor perturbations are generated by inflation, as well as in any other model in which the   primordial tensor perturbations  are not completely negligible, the presence of the non-local Weyl-square term in the effective action at late time is ruled out by the resulting overproduction of tensor perturbations.

\section{Conclusions}\label{sect:concl}

The aim of this paper was to study a non-local theory of the general form (\ref{actionTotal}), which features non-local terms proportional to the mass-square parameters $\mu_1,\mu_2,\mu_3$,
to see which terms are phenomenologically allowed. The term $\mu_1 R\Box^{-2}R$ has already been studied in a number of papers, and has been shown to produce a viable cosmological model, both at the level of background evolution and of cosmological perturbations.
Here we have found that the addition of a term $\mu_3\Rmn\Box^{-2}\RMN$ is quickly ruled out, since it induces instabilities in the  background evolution. The case for the non-local Weyl-square term, 
$\mu_2 C^{\mu\nu\rho\sigma}\Box^{-2}C_{\mu\nu\rho\sigma}$, turns out to be more delicate. This term does not contribute to the background evolution. Its scalar perturbations, at least up to the present epoch, are sufficiently close to that of {\LCDM} and of the 
$\mu_1 R\Box^{-2}R$ model, so that they do not pose a threatens to the phenomenology of the model (although, for very large wavelengths, they become unstable in the cosmological future). However, the tensor perturbations of the model become unstable already during MD, with the instability being more important for the short wavelengths. This produces a strong amplification of any primordial background of gravitational waves. In particular, if we assume the typical primordial GW background generated by single-field slow-roll inflation, even with very small values of the tensor-to-scalar ratio $r$, the resulting spectrum for $\hogw(f)$ is extremely blue, growing as $f^4$ for modes that re-entered in RD, leading to a complete violation of all bounds on $\hogw(f)$. Indeed, increasing the frequency, $\hogw(f)$ even quickly becomes larger than one. Thus, at least in the standard scenario where the initial relic GW background is not totally negligible, also the term 
$\mu_2 C^{\mu\nu\rho\sigma}\Box^{-2}C_{\mu\nu\rho\sigma}$ is phenomenologically ruled out.\footnote{We have found a similar instability in the vector sector of the Weyl-square model, although in this case inflation can dilute to totally negligible value any initial primordial spectrum.}

The crucial next step is  to understand the mechanism that generates such non-localities from a fundamental local theory. The results that we have presented in this paper give a useful hint, since they indicate that we need a mechanism that generates precisely the $R\Box^{-2}R$ term (or the term in \eq{RT}), rather than  most general non-local structures involving $\Rmn$ ot the Weyl tensor.  A mechanism of this type has been recently suggested in \cite{Maggiore:2015rma}. 
The starting point was the observation in \cite{Antoniadis:1991fa,Antoniadis:1996pb} that, when gravity is coupled to massless particles, the conformal mode acquires a kinetic term because of the conformal anomaly. In four dimensions the anomaly-induced effective action is
\be\label{Sanom}
S_{\rm anom}=-\frac{1}{8}\int d^4x\sqrt{-g}\(E-\frac{2}{3}\Box R\) \Delta_4^{-1}
\[ b' \(E-\frac{2}{3}\Box R\)-2b C^2\]\, ,
\ee
where $E$ is the Gauss-Bonnet combination, $C^2$ is the square of the Weyl tensor, $b,b'$ are coefficients that depend on the number and type of conformal massless fields, and $\Delta_4$ was defined in \eq{Delta4}.
Restricting the theory to the conformal mode $\sigma$, i.e. writing the metric as $\gmn(x)=e^{2\sigma(x)}$, the non-local action (\ref{Sanom}) becomes local, and reads \cite{Antoniadis:1991fa,Antoniadis:2006wq}
\be\label{Sanom4D}
S_{\rm anom}=-\frac{Q^2}{16\pi^2}\int d^4x\, (\Box\sigma)^2\, ,
\ee
where again $Q$  is a coefficient that depends on the number and type of conformal massless fields. The interesting aspect of this result is that, in momentum space, the propagator of the $\sigma$ field goes as $1/k^4$, and therefore has strong infrared (IR) effects. In particular, the corresponding propagator in coordinate space grows logarithmically,
$G(x,x')=-(2Q^2)^{-1}\log[\mu^2 (x-x')^2]$. The situation is quite similar to what happens in  the two-dimensional XY model. In this case the $1/k^2$ propagator in two dimensions again produces a coordinate-space propagator which grows logarithmically. In the XY model this induces  a vortex-vortex interaction that, depending on the value of $Q$, can disorder the system and dynamically generate a mass gap, leading to the well-known 
Berezinsky-Kosterlitz-Thouless (BKT) transition. It is then quite natural to expect that, in the four-dimensional case, the strong IR effects induced by the $1/k^4$ propagator of the conformal mode might lead to a dynamical mass generation for the conformal mode, similarly to the dynamical mass generation in the BKT transition. In this case, it is precisely the term  $m^2R\Box^{-2}R$ that emerges~\cite{Maggiore:2015rma}. Indeed, there is no local diff-invariant term that, when written in terms of the conformal mode and expanded in powers of $\sigma$, gives a mass term for $\sigma$, i.e. starts from a $\sigma^2$ term. However,
the Ricci scalar computed from the metric $\gmn=e^{2\sigma(x)}\emn$, to linear order in $\sigma$, is given by
$R=-6\Box\sigma +{\cal O}(\sigma^2)$. Thus,
 (upon integration by parts)
\be\label{m2s2}
 m^2R\frac{1}{\Box^2} R=36m^2 \sigma^2 +{\cal O}(\sigma^3)\, .
\ee
The  non local term  $m^2R\Box^{-2}R$   therefore just provides a diff-invariant way of giving a mass  to the conformal mode, while the higher-order interaction terms on the right-hand side of \eq{m2s2} (which are non-local even in $\sigma$) are required to reconstruct a diff-invariant quantity. The same is true for the model  (\ref{RT}). Indeed, as we have mentioned, the models (\ref{RT})  and (\ref{RR}) have the same expansion
 to linear order over Minkowski space, so the corresponding actions are the same in an expansion to  quadratic order over Minkowski, and they both reproduce the term $\propto\sigma^2 $ in the action, although with different non-linear completions. 

It is quite interesting to see that the models  (\ref{RT})  and (\ref{RR}) have a physical interpretation, as a mass term for the conformal mode, which is absent for the other non-local terms that we have considered, and which could justify the emergence of this specific non-local structure. Further work along these lines is of course needed to put this picture on firmer grounds. The results of the present paper however  show that, among the possible non-local terms of the form (\ref{actionTotal}), only the 
$R\Box^{-2}R$ term is phenomenologically acceptable, and this points toward a mechanism, such as the dynamical mass generation for the conformal mode, which would produce exactly this non-local structure.

\vspace{5mm}
\noindent
{\bf Acknowledgments.}
The work of GC, SF and MM is supported by the Fonds National Suisse. The work of MM is supported by the SwissMap NCCR.
\appendix

\section{Scalar perturbations with $k/H_0\ll 1$}\label{sect:appSH}

In this appendix we compute analytically the behavior of scalar perturbations with very low values of $k$, e.g. $k/H_0=10^{-3}$. As we see from Fig.~\ref{fig:scalars}, contrary to what happens in {\LCDM}, these modes eventually become non-linear in the cosmological future. This behavior can be understood analytically, as follows.

\begin{figure}[ht!]
   \vspace{1 em}
  \centering
  {\includegraphics[scale=0.25]{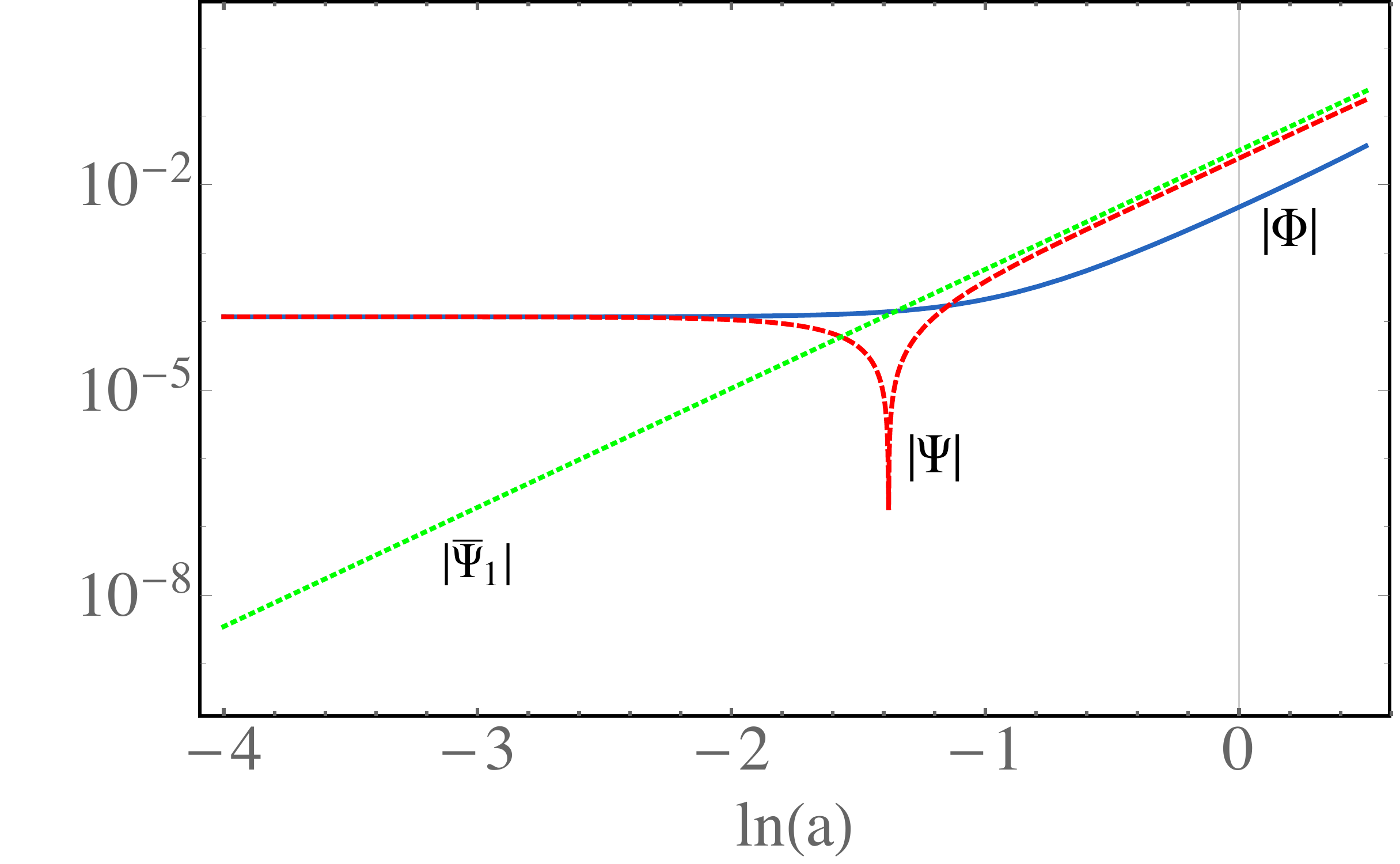}}\qquad
  {\includegraphics[scale=0.25]{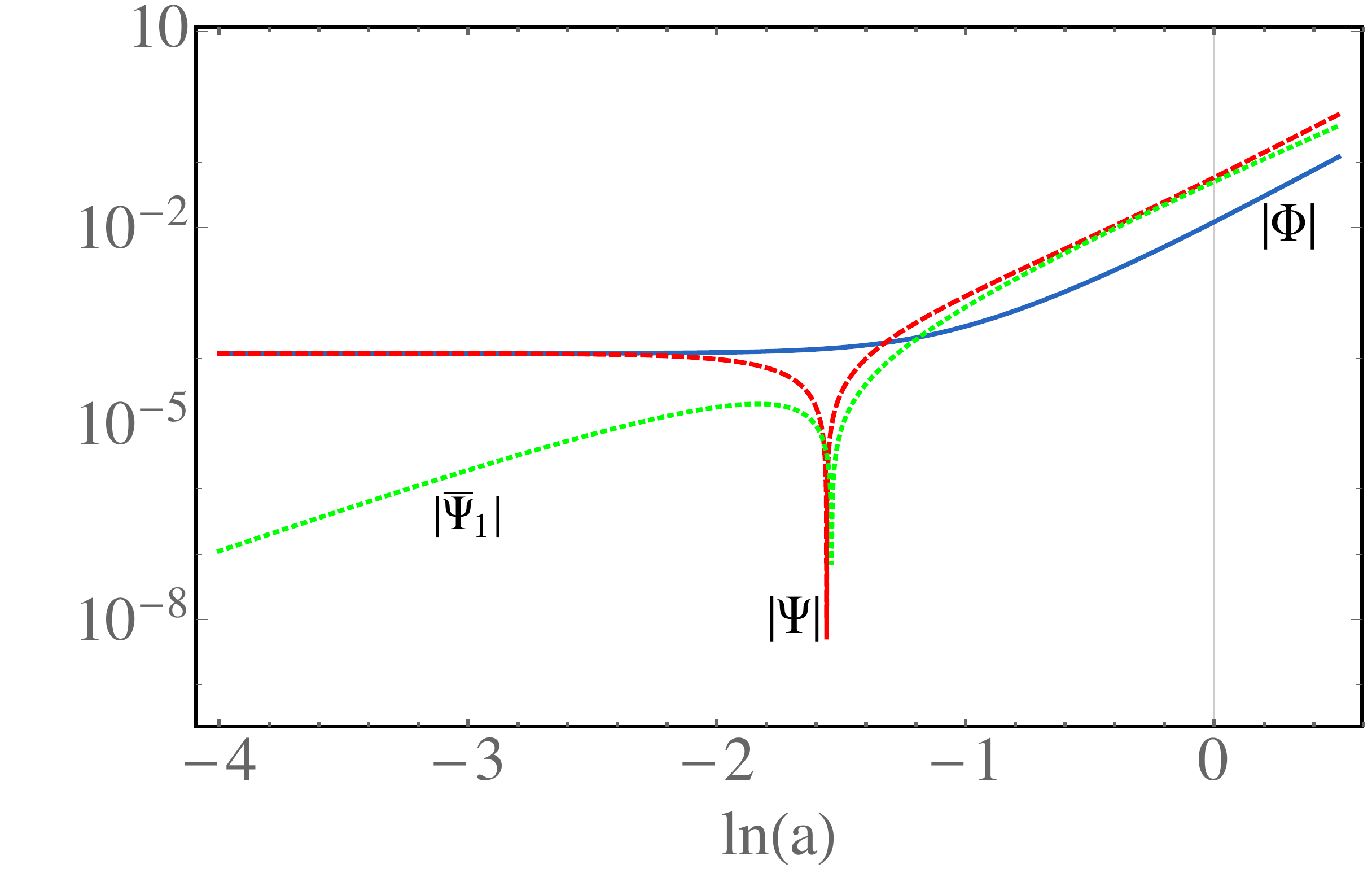}}
 \caption{\label{fig1scalar} \small Numerical evolution of the potentials $\Phi$ (blue, solid line) and $\Psi$ (red, dashed line), with $k=10^{-3}\mathcal{H}_0$ and $\gamma_2=\gamma_1=0.014 H_0^2$. Left: the background evolution is driven by standard matter and radiation only. The analytical solution for $\Psi$ in matter dominance is the green, dotted line.  Right: the background evolution includes dark energy. When the latter becomes relevant at the background level, the solution follows the deSitter growing mode. 
  }
 \end{figure}

The evolution of scalar perturbations is governed by  \eqst{eqs1}{eqs6} and  (\ref{eqs7})--(\ref{eqs10}).
These equations simplify for  modes with $k\ll  aH$, and the possible onset of an instability can be studied 
analytically setting $\gamma_1=0$ in the equations for the perturbations (since the perturbations of the $R\Box^{-2}R$ model do not show this instability) and working perturbatively in $\gamma_2$. Observe that the background evolution is still computed using a non-vanishing $\gamma_1$. We consider first the behavior in RD and MD. 
During radiation and matter dominance, the function $\zeta$ can be approximated as a constant, $\zeta_0=\{-2,-3/2\}$ respectively.
At zeroth order in  $\gamma_2$ and for $k\ll  aH$, \eqst{eqs1}{eqs2} have the GR solution (in both radiation and matter dominance)
\be\label{Phi0const}
\Psi_{0}=-\Phi_{0}=\text{const.}\;.
\ee
Plugging this solution into equation (\ref{eqs6}), we can solve for the auxiliary field $u$ and the $ s$ at zeroth order, 
and we find that, both in RD and MD,
the fastest growing mode are given by the particular solution, $u_{0}\propto a^2/h^2$ and  $s_{0}\propto a^2/h^4$.
The zeroth-order evolution of the auxiliary fields $u,s$ acts as a source for the first-order correction to the gravitational potentials in \eqst{eqs1}{eqs2}, which to this order become
\bees
&&\Phi_1''+\left(\zeta+3\right)\Phi_1'-\Psi_1 \left(3+2\zeta\right)
-\Psi_1'= 0  \label{phisec1st}\\ 
&&\Psi_1=-\Phi_1-\frac{6\,\gamma_2}{a^2}h^2\,\left[{\hat{s}}_0''+\left(\zeta-3\right){\hat{s}}_0'-\left(\frac{5}{4}\zeta-2\right)\hat{s}_0+\frac{\hat{k}^2}{3}\hat{s}_0\right]\label{phipsi1st} 
\ees
The most relevant correction comes from the effective anisotropic stress generated by the effective dark energy in equation (\ref{eqs2}), which makes $\Phi \neq \Psi$. The other contributions in equation  (\ref{eqs1}) are suppressed by extra factors of $\hat{k}^2$ and have been neglected in (\ref{phisec1st}).
 Note that we must however retain the term in $\hat{k}^2$ in (\ref{eqs2}) since it may become relevant at late times: indeed, taking for definiteness a MD epoch ($h \propto a^{-3/2}$), and remembering that $\hat{k}=k/(a h) $, we see that the contribution to the effective anisotropic stress proportional to $k$ on the right-hand side of \ref{phipsi1st} is diluted by a factor $a^{-4}$, while the other source terms are diluted by $h^2/a^2 \propto a^{-5} $.
 We can solve algebraically (\ref{phipsi1st}) for $\Psi_1$ and use the solution to get a second-order differential equation for $\Phi_1$ from (\ref{phisec1st}).
Then, to first order in $\gamma_2$, the particular solution for the gravitational potential $\Phi$ in MD contains a growing mode, 
\be \label{phi1grsol}
\Phi_1 \propto k^2 a/H^2 \propto a^4 \; , 
\ee
 which, at late times, dominates over the constant solution. 
In the left panel of Fig.~\ref{fig1scalar} we show the numerical  evolution of the potentials $\Phi$ and $\Psi$ in  MD, in the approximation in which
the effect of dark energy on the background evolution is neglected (an approximation that, as we know from Sect.~\ref{sect:back}, breaks down near the present epoch). In this case, at late times the universe is dominated by matter and the potentials switches from the constant behavior (\ref{Phi0const}) at the beginning of the MD epoch, to the analytic behavior (\ref{phi1grsol}) (red, dashed line) induced by the growing mode at later times.

We next consider the dark-energy dominated regime. We mimick it with a
de Sitter phase with $\zeta=0$, while $h^2$ and $\Omega_{\rm DE}$ are taken to be constant; we will see that this rather crude approximation (our nonlocal effective dark energy fluid does not behave as a cosmological constant) captures anyways the essential behaviour of superhorizon scales. Then, the growing modes of the auxiliary fields  become dominated by the  solutions $u\propto a^{\lambda_{+}}/h^2$, $s\propto a^{\lambda_{+}}/h^4$, with $\lambda_{+}=(5+\sqrt{33})/2\approx 5.37$.
We can then proceed as before and compute the correction to the scalar potentials, which now has a growing mode $\Phi \propto (\alpha + \beta x) a^{\lambda_{+} x}/h^2$. In the right panel of Fig. \ref{fig1scalar}, we show the evolution of the potentials on the background given by the full non-local model. At late times, the growing mode of the solution in de Sitter becomes dominating. Because of this growing behavior,  scalar perturbations of very large wavelength eventually leave the linear regime, even if this happens only in the cosmological future. We see that these  analytic estimates allow us to correctly reproduce the qualitative features of the numerical results.

\section{Analytic study of the instabilities in the tensor sector}\label{App:tensor} 

In this section we show how to understand analytically some aspects of the evolution of  the tensor perturbations, studied numerically in Sect.~\ref{sect:tenssect}.

We consider first  superhorizon scales. Again, we solve first the equations to zero-th order in $\gamma_1$, keeping only the leading growing terms, and we then plug the solution back into the equation for $h_g$. 
To first order  in $\gamma_2$ we find
\be\label{hrad}
h''_g+h'_g+\hat{k}^2 h_g=\frac{6\gamma_2}{a^4}\left[2s' _g-2\left(\frac{k}{H_0}\right)^2\frac{a^2}{\Omega_R}\,s_g-\frac{a^4}{\Omega_R}\,u_g-2 a\, \Omega_R^{-1/2}\,\sigma'_g\right]\,.
\ee
The tensor mode $h_g$ will experience the effects of the source term on the right-hand side of \eq{hrad} when the latter will have grown of order of the $\hat{k}^2 h_g$ term present on the left-hand side.  Substituting the solutions 
for $u_g$,$v_g$ $s_g$ and $\sigma_g$ we find that the leading-order growth of the source term is given by $\sim \gamma_2\,\left(k/H_0\right)^4 \Omega_R^{-3}a^8\log a$. Therefore, a mode with momentum $k$ will show an instability only  for values of the redshift that satisfy
$z^6 \leq \Omega_R^{-2} \gamma_2\,(k/H_0)^2$. 
On the other hand, the super-horizon approximation is valid for scales satisfying the  condition 
$z^6\gg (k/H_0)^6\,\Omega_R^{-3}$. These two
conditions cannot  be satisfied at the same time, unless an unnaturally large value of $\gamma_2$ is chosen. Therefore, modes that enter the horizon after the equality are stable during the entire radiation dominated epoch. For these modes the effect of the coupling in eq.~(\ref{hrad}) is negligible and their evolution is the standard one in term of a constant and a decaying mode. This agrees indeed with our numerical result for the mode with $\lbar =1$~Gpc, shown in Fig.~\ref{fig:tensorLog}.

\begin{figure}[t]
   \vspace{1 em}
  \centering
  {\includegraphics[scale=0.35]{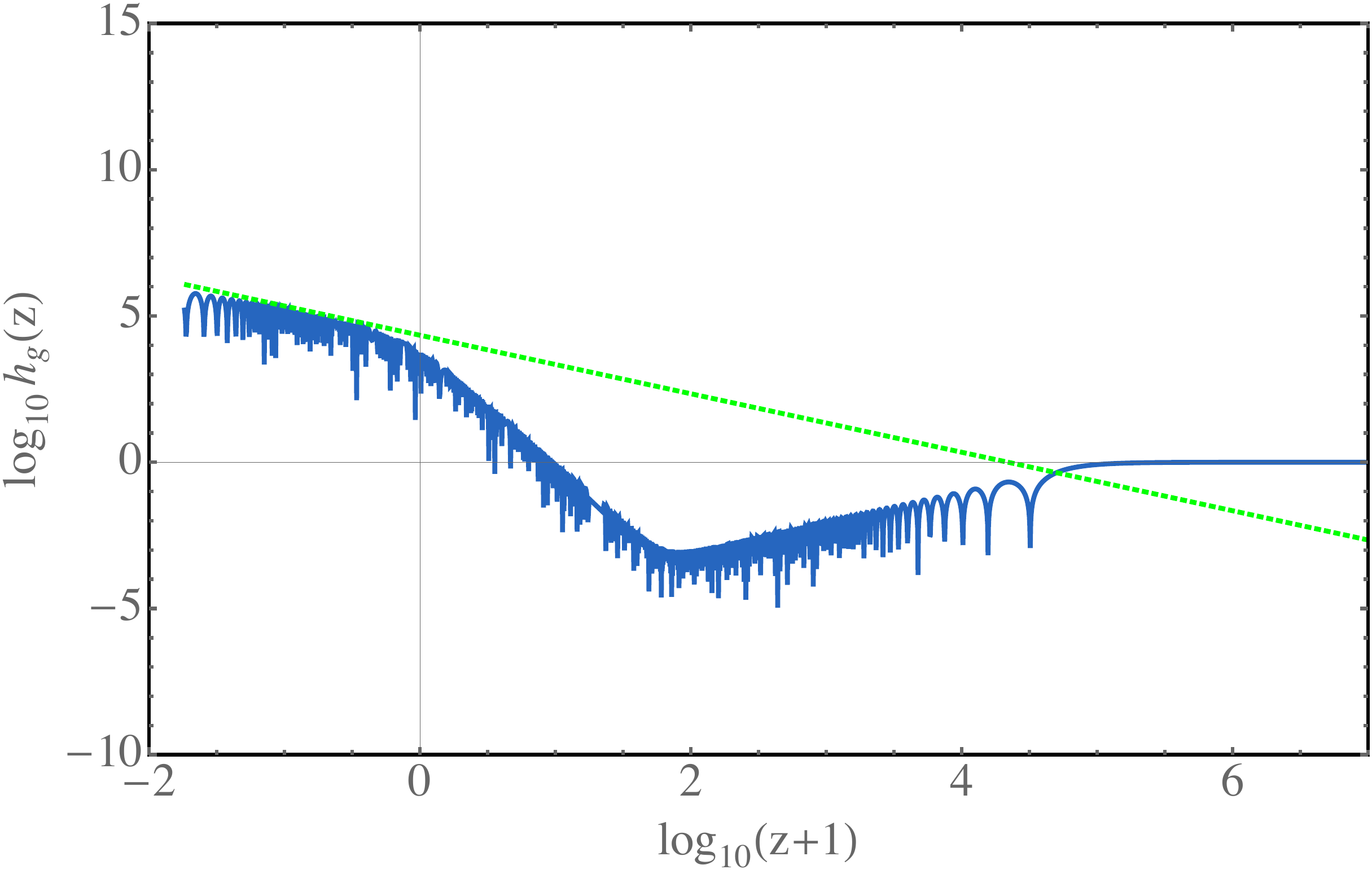}}\qquad
 \caption{\label{fig2} \small Numerical evolution of the amplitude of the metric tensor mode $h_g$ for $k=10^3 H_0$ and $\gamma_2=\gamma_1=0.014 H_0$. The numerical solution (blue, solid line) is compared to the analytic one valid for sub-horizon modes at late times (green, dotted line).}
 \end{figure}

We next consider the late-time instability of sub-horizon modes.
A full analytic understanding of the growth of sub-horizon perturbations in the different epochs  is rather complicated since several terms in the solution grow with different powers of the scale factors, but also with different $\hatk$-dependent pre-factors. Then, it is quite involved to see, as a function of $\hatk$, what is the term that dominates the growth at any given time. However, it is relatively straightforward to understand the behavior of sub-horizon modes at late times, i.e. deep in the DE dominated epoch (so, in particular, in the cosmological future), since in this case the result is dominated by the terms that grows with the highest power of $a$.
As for the scalar case, we approximate the late phase of the evolution of the background as a pure de~Sitter phase  with $\zeta=0$ and $h=\Omega_{\rm DE}=$~constant. As one can check self-consistently, the fastest-growing modes are the ones associated to the homogeneous solutions of eqs. (\ref{stens}), (\ref{sigmatens}),  (\ref{utens}) and (\ref{vtens}).  These equations have a leading-order  growth given by
$u_g\,,v_g\, ,s_g\,,\sigma_g \propto a^5$.
Making use  of this behavior for the  auxiliary fields, to first order in $\gamma_2$ the eqaution for $h_g$  can schematically be  written as
\be\label{hde}
h''_g+4h'_g+\hat{k}^2 h_g-\gamma_2 \left(c_1 a^{-3}+c_2 a^{-1}+c_3 a\right)=0\,,
\ee
where $c_1,c_2$ and $c_3$ are some coefficients constant  in time, but which depend on $\hatk$, as well as on $\Omega_R$, $\Omega_M$, $\Omega_{\rm DE}$ and $z_i$. At late times, the GW mode $h_g$ will then develop a growing mode, with leading order behavior 
\be\label{hgpropa}
h_g\propto a\,.
\ee
In Fig.~\ref{fig2} we plot the numerical evolution of a tensor mode with very short wavelength, $k=10^3 H_0$,  
and we compare it to the slope of the analytic estimate (\ref{hgpropa}) in the DE dominated epoch (observe that we now plot the result as a function 
of $\log_{10}(z+1)$, where $z$ is the red-shift, so time evolves from right to left).
We see that the analytic estimate correctly reproduces the numerical result at late times.

\bibliographystyle{utphys}
\bibliography{myrefs_massive}

\end{document}

%% file: mydefs.tex
\newcommand{\nn}{\nonumber}

\newcommand{\iBox}{\Box^{-1}}

\renewcommand\({\left(}
\renewcommand\){\right)}
\renewcommand\[{\left[}
\renewcommand\]{\right]}

\newcommand\n{{\mbox {\boldmath $\nabla$}}}

\def\lsim{\raise 0.4ex\hbox{$<$}\kern -0.8em\lower 0.62
ex\hbox{$\sim$}}

\def\gsim{\raise 0.4ex\hbox{$>$}\kern -0.7em\lower 0.62
ex\hbox{$\sim$}}

\def\lbar{{\hbox{$\lambda$}\kern -0.7em\raise 0.6ex
\hbox{$-$}}}

\newcommand\eq[1]{eq.~(\ref{#1})}
\newcommand\eqs[2]{eqs.~(\ref{#1}) and (\ref{#2})}

\newcommand\eqst[2]{eqs.~(\ref{#1})--(\ref{#2})}

\newcommand\pa{\partial}
\newcommand\p{\partial}

\newcommand\ee{\end{equation}}
\newcommand\be{\begin{equation}}
\def\bea{\begin{array}}
\def\eea{\end{array}}\def\ea{\end{array}}
\newcommand\ees{\end{eqnarray}}
\newcommand\bees{\begin{eqnarray}}
\def\nn{\nonumber}

\def\d{\delta}

\def\dslash{\hspace{-1mm}\not{\hbox{\kern-2pt $\partial$}}}
\def\Dslash{\not{\hbox{\kern-4pt $D$}}}
\def\pslash{\not{\hbox{\kern-2.1pt $p$}}}
\def\kslash{\not{\hbox{\kern-2.3pt $k$}}}
\def\qslash{\not{\hbox{\kern-2.3pt $q$}}}

\newcommand{\vk}{{\bf k}}
\newcommand{\vx}{{\bf x}}

\def\p1{{\bf p}_1}
\def\p2{{\bf p}_2}
\def\k1{{\bf k}_1}
\def\k2{{\bf k}_2}

\newcommand{\emn}{\eta_{\mu\nu}}

\newcommand{\gmn}{g_{\mu\nu}}

\newcommand{\hatk}{\hat{\bf k}}

\newcommand{\pam}{\pa_{\mu}}

\newcommand{\pan}{\pa_{\nu}}
\newcommand{\parho}{\pa_{\rho}}

\newcommand{\paR}{\pa^{\rho}}

\newcommand{\Rmn}{R_{\mu\nu}}
\newcommand{\Gmn}{G_{\mu\nu}}
\newcommand{\RMN}{R^{\mu\nu}}

\newcommand{\Rmnrs}{R_{\mu\nu\rho\sigma}}
\newcommand{\RMNRS}{R^{\mu\nu\rho\sigma}}

\newcommand{\Tmn}{T_{\mu\nu}}

\newcommand{\dddM}{\kern 0.2em \raise 1.9ex\hbox{$...$}\kern -1.0em \hbox{$M$}}
\newcommand{\dddQ}{\kern 0.2em \raise 1.9ex\hbox{$...$}\kern -1.0em \hbox{$Q$}}
\newcommand{\dddI}{\kern 0.2em \raise 1.9ex\hbox{$...$}\kern -1.0em\hbox{$I$}}
\newcommand{\dddJ}{\kern 0.2em \raise 1.9ex\hbox{$...$}\kern-1.0em
\hbox{$J$}}
\newcommand{\dddcalJ}{\kern 0.2em \raise 1.9ex\hbox{$...$}\kern-1.0em
\hbox{${\cal J}$}}

\newcommand{\dddO}{\kern 0.2em \raise 1.9ex\hbox{$...$}\kern -1.0em
\hbox{${\cal O}$}}
\def\dddz{\raise 1.5ex\hbox{$...$}\kern -0.8em \hbox{$z$}}
\def\dddd{\raise 1.8ex\hbox{$...$}\kern -0.8em \hbox{$d$}}
\def\dddbd{\raise 1.8ex\hbox{$...$}\kern -0.8em \hbox{${\bf d}$}}
\def\ddbd{\raise 1.8ex\hbox{$..$}\kern -0.8em \hbox{${\bf d}$}}
\def\dddx{\raise 1.6ex\hbox{$...$}\kern -0.8em \hbox{$x$}}

\newcommand{\mplr}{m_{\rm Pl}}

\newcommand{\hogw}{h_0^2\Omega_{\rm gw}}

\newcommand{\ode}{\Omega_{\rm DE}}
\newcommand{\oma}{\Omega_{M}}

\newcommand{\ola}{\Omega_{\Lambda}}